\documentclass[11pt,a4paper]{article}
\pdfoutput=1
\usepackage{jheppub}

\usepackage{placeins}
\usepackage{epsfig}
\usepackage{latexsym}
\usepackage[bbgreekl]{mathbbol}
\usepackage{mathtools}
\usepackage{eqnarray,amsmath,amsfonts,amstext,amssymb,mathrsfs}
\usepackage{multirow}
\usepackage{slashed}
\usepackage{float}
\usepackage{esint}
\usepackage{epsfig}
\usepackage{lscape}
\usepackage{graphicx}
\usepackage{psfrag}
\usepackage[utf8]{inputenc}
\usepackage{subfigure}
\usepackage{nicefrac}
\usepackage[stable]{footmisc}
\usepackage{rotating}
\usepackage{afterpage}
\usepackage{bbm}
\usepackage{color,xcolor}
\usepackage{cancel}

\def\){\right)}
\def\({\left( }
\def\]{\right] }
\def\[{\left[ }

\def\NO{\nonumber}

\newcommand{\be}{\begin{equation}}
\newcommand{\ee}{\end{equation}}

\def\bea{\begin{eqnarray}}
\def\eea{\end{eqnarray}}

\def\bal#1\eal{\begin{align}#1\end{align}}

\def\bald{\begin{aligned}}
\def\eald{\end{aligned}}

\def\bsub{\begin{subequations}}
\def\esub{\end{subequations}}

\def\beqx{\begin{displaymath}}
\def\eeqx{\end{displaymath}}

\newcommand{\bmat}{\left(\begin{array}}
\newcommand{\emat}{\end{array}\right)}

\def\a{\alpha}

\def\c{\chi}
\def\d{\delta}
\def\e{\epsilon}
\def\f{\phi}
\def\g{\gamma}
\def\h{\eta}
\def\j{\psi}
\def\k{\kappa}
\def\l{\lambda}
\def\m{\mu}
\def\n{\nu}
\def\o{\omega}
    
\def\p{\pi}

    \def\th{\theta}
\def\r{\rho}
\def\s{\sigma}
\def\t{\tau}
\def\x{\xi}

\def\G{\Gamma}
\def\J{\Psi}
\def\L{\Lambda}
\def\O{\Omega}

    \def\Th{\Theta}
\def\S{\Sigma}

\def\X{\Xi}

\def\ve{\varepsilon}

\def\ca{{\cal A}}

\def\cd{{\cal D}}

\def\cj{{\cal J}}
\def\ck{{\cal K}}

\def\cn{{\cal N}}
\def\co{{\cal O}}
\def\cp{{\cal P}}
\def\cq{{\cal Q}}

\def\cs{{\cal S}}
\def\ct{{\cal T}}

\def\bo{{\raise-.3ex\hbox{\large$\Box$}}}               
\def\pa{\partial}                                       
\def\de{\nabla}                                         
\def\face{{\raise.2ex\hbox{$\displaystyle \bigodot$}\mskip-2.2mu \llap {$\ddot
        \smile$}}}                                   
\def\>{\rangle}                                      
\def\<{\langle}                                      

\def\tx#1{\text{#1}}
\def\wt#1{\widetilde{#1}}                            
\def\Hat#1{\widehat{#1}}                             
\def\lbar#1{\ensuremath{\overline{#1}}}              
\def\leftrightarrowfill{$\mathsurround=0pt \mathord\leftarrow \mkern-6mu
        \cleaders\hbox{$\mkern-2mu \mathord- \mkern-2mu$}\hfill
        \mkern-6mu \mathord\rightarrow$}        
\def\dvec#1{\vbox{\ialign{##\crcr
        \leftrightarrowfill\crcr\noalign{\kern-1pt\nointerlineskip}
        $\hfil\displaystyle{#1}\hfil$\crcr}}}           
\def\Tr{{\rm Tr \,}}                                    

\def\-{\hphantom{-}}

\title{Supersymmetry anomalies in new minimal supergravity}
\author[a]{Ioannis Papadimitriou} 

\affiliation[a]{School of Physics, Korea Institute for Advanced Study, 85 Hoegiro, Seoul 02455, Korea}

\emailAdd{ioannis@kias.re.kr}

\abstract{We determine the general structure of quantum anomalies for the $R$-multiplet of four dimensional $\cn=1$ supersymmetric quantum field theories in the presence of background fields for an arbitrary number of Abelian flavor multiplets. By solving the Wess-Zumino consistency conditions for off-shell new minimal supergravity in four dimensions with an arbitrary number of Abelian vector multiplets, we compute the anomaly in the conservation of the supercurrent to leading non trivial order in the gravitino and vector multiplet fermions. We find that both $R$-symmetry and flavor anomalies necessarily lead to a supersymmetry anomaly, thus generalizing our earlier results to non superconformal theories with Abelian flavor symmetries. The anomaly in the conservation of the supercurrent leads to an anomalous transformation for the supercurrent under rigid supersymmetry on bosonic backgrounds that admit new minimal Killing spinors. The resulting deformation of the supersymmetry algebra has implications for supersymmetric localization computations on such backgrounds.}


\preprint{KIAS-P19022}

\begin{document}  
\maketitle



\section{Introduction}
\label{intro}

Supersymmetric quantum field theories are an invaluable tool for probing strong coupling dynamics. Unbroken supersymmetry permits the use of non renormalization theorems and supersymmetric localization techniques \cite{Witten:1982im,Witten:1988ze,Nekrasov:2002qd,Pestun:2007rz} in order to non perturbatively compute observables such as partition functions and Wilson loops. Supersymmetry is also relevant for extending the Standard Model to higher energies and plays a pivotal role in holographic dualities and string theory. A question of paramount importance, therefore, is whether supersymmetry is anomalous at the quantum level.  

Several supersymmetry anomalies have been discussed in the literature and fall into two broad classes, depending on whether they appear in the gamma trace or in the divergence of the supercurrent. The gamma trace of the supercurrent is in the same multiplet as the trace of the stress tensor and the divergence of the $R$-current \cite{Ferrara:1974pz} and so the corresponding supersymmetry anomalies are part of the multiplet of superconformal anomalies \cite{McArthur:1983fk,Bonora:1984pn,Buchbinder:1986im,Buchbinder:1988yu,Brandt:1993vd,Brandt:1996au,Anselmi:1997am,Piguet:1998bj,Erdmenger:1998ew,Bonora:2013rta,Butter:2013ura,Cassani:2013dba,Auzzi:2015yia}. Anomalies in the gamma trace of the supercurrent arise also in non Abelian supersymmetric gauge theories if one insists on a gauge invariant supercurrent that is conserved \cite{deWit:1975veh,Abbott:1977in,Abbott:1977xj,Abbott:1977xk,Hieda:2017sqq,Batista:2018zxf}.

The supersymmetry anomalies we are concerned with here, however, are those arising in the divergence of the supercurrent. Such anomalies have been less studied and are often believed to be absent in physical theories. The first examples of supersymmetry anomalies in the divergence of the supercurrent were found in the context of supersymmetric theories with gauge anomalies. In particular, the fact that the Wess-Zumino consistency conditions \cite{Wess:1971yu} imply the presence of a supersymmetry anomaly whenever the theory has a gauge anomaly was pointed out in \cite{Itoyama:1985qi} (see also \cite{Piguet:1984aa,Guadagnini:1985ea, Zumino:1985vr} and \cite{Piguet:1986ug} for a review). However, gauge anomalies must be canceled for the consistency of the theory at the quantum level, and so the corresponding supersymmetry anomaly is canceled as well. An anomaly in the divergence of the supercurrent was also found in the presence of a gravitational anomaly in two-dimensional theories in \cite{Howe:1985uy,Itoyama:1985ni,Tanii:1985wy}. This anomaly is conceptually closer to the supersymmetry anomalies we discuss here since it is related to a global anomaly, which need not be canceled. 

Anomalies in the divergence of the supercurrent have also been discussed in the context of supergravity theories \cite{Shamir:1992ff,Brandt:1993vd,Brandt:1996au}. These works focused on dynamical or on-shell supergravity, but some of the supersymmetry anomalies identified there appear as well in off-shell background supergravity, which is relevant for studying {\em global} supersymmetry anomalies in supersymmetric quantum field theories. Global anomalies are a property of the theory and do not lead to any inconsistencies. They have physical consequences, such as the violation of selection rules \cite{Adler:1969gk,Bell:1969ts} and the transport properties of the theory \cite{Landsteiner:2016led}. In particular, global supersymmetry anomalies do not render a quantum field theory inconsistent, but they imply that supersymmetry cannot be gauged, i.e. the theory cannot be consistently coupled to dynamical supergravity at the quantum level. Moreover, global supersymmetry anomalies may violate some of the conditions required in order for non-renormalization theorems and supersymmetric localization techniques to be applicable.  

Classifying global supersymmetry anomalies is therefore particularly relevant  
following the recent advances in supersymmetric localization techniques for quantum field theories on curved backgrounds \cite{Pestun:2007rz} (see \cite{Pestun:2016zxk} for a comprehensive review). A systematic way for placing supersymmetric quantum field theories on curved backgrounds was developed in \cite{Festuccia:2011ws} and involves coupling the theory to a given off-shell background supergravity. This corresponds to turning on background fields for the current multiplet operators. Rigid supersymmetry on purely bosonic backgrounds can then be defined independently of the details of the microscopic theory through the Killing spinor equations obtained by setting to zero the supersymmetry variations of the background supergravity fermions. This procedure leads to a classification of supersymmetric backgrounds preserving a number of supercharges
\cite{Samtleben:2012gy,Klare:2012gn,Dumitrescu:2012ha,Liu:2012bi,Dumitrescu:2012at,Kehagias:2012fh,Closset:2012ru,Samtleben:2012ua,Cassani:2012ri,deMedeiros:2012sb,Kuzenko:2012vd,Hristov:2013spa,Pan:2013uoa,Imamura:2014ima,Alday:2015lta} (see also \cite{Blau:2000xg} for earlier work). However, the corresponding rigid supersymmetry may or may not be preserved at the quantum level. 

The fact that rigid supersymmetry defined in this way can be anomalous at the quantum level was first pointed out in the context of theories with a holographic dual \cite{Papadimitriou:2017kzw,An:2017ihs,An:2018roi}. The anomaly in rigid supersymmetry refers to a local term in the quantum supersymmetry transformation of the supercurrent and depends on the bosonic background. This bosonic term is directly related to the (fermionic) supersymmetry anomalies in the divergence and (in the case of conformal supergravity backgrounds) the gamma trace of the supercurrent. The form of these anomalies for any four dimensional superconformal field theory on backgrounds of $\cn=1$ conformal supergravity was derived in \cite{Papadimitriou:2019gel} by solving the corresponding Wess-Zumino consistency conditions. The presence of these supersymmetry anomalies was also verified through a perturbative calculation of flat space four-point functions involving two supercurrents and either two $R$-currents or one $R$-current and a stress tensor in the free and massless Wess-Zumino model \cite{Katsianis:2019hhg,followup}.

In this paper we consider off-shell new minimal supergravity in four dimensions  \cite{Akulov:1976ck,Sohnius:1981tp,Sohnius:1982fw,Ferrara:1988qxa} in the presence of an arbitrary number of Abelian vector multiplets. This provides a suitable set of background fields for the $R$-multiplet of current operators that exists for supersymmetric theories with a U(1) $R$-symmetry \cite{Gates:1983nr,Komargodski:2010rb}, as well as for an arbitrary number of flavor multiplets. We determine the algebra of local symmetry transformations and identify a specific relation with the symmetry algebra of $\cn=1$ conformal supergravity. This allows us to derive the supersymmetry anomalies of the new minimal gravity multiplet from those of $\cn=1$ conformal supergravity obtained in \cite{Papadimitriou:2019gel}. Six additional candidate anomalies are found in the presence of vector multiplets by directly solving the Wess-Zumino consistency conditions for new minimal supergravity to leading non trivial order in the gravitino and the flavorinos. These results extend our earlier analysis for $\cn=1$ conformal supergravity \cite{Katsianis:2019hhg,Papadimitriou:2019gel} to non conformal theories with an arbitrary number of Abelian flavor symmetries. We find that the presence of either an $R$-symmetry or a flavor symmetry anomaly necessarily leads to a supersymmetry anomaly, irrespective of whether the theory is conformal or not. This result is consistent with the observation of \cite{An:2019zok} that in theories with an $R$-multiplet supersymmetry can be non anomalous provided $R$-symmetry is non anomalous.  
The supersymmetry anomaly is cohomologically non trivial and cannot be removed by a local counterterm without breaking diffeomorphism and/or local Lorentz symmetry. Moreover, it implies that the fermionic operators in the current and flavor multiplets acquire an anomalous supersymmetry transformation at the quantum level, even on purely bosonic backgrounds. The significance of this anomalous transformation for supersymmetric quantum field theory observables on new minimal supergravity backgrounds that preserve a number of supercharges is discussed. 

The paper is organized as follows. In section \ref{nm-algebra} we review the local symmetry algebra of off-shell new minimal supergravity in four dimensions and we discuss its relation to the symmetry algebra of $\cn=1$ conformal supergravity. In section \ref{nm-anomalies} we utilize this relation in order to derive the Ward identities and anomaly candidates for the gravity multiplet of new minimal supergravity from those of $\cn=1$ conformal supergravity. These results are generalized in section \ref{nm-vector-multiplet} to include background fields for an arbitrary number of Abelian flavor multiplets. In section \ref{supercurrent} we derive the anomalous supersymmetry transformations of the supercurrent and of the fermionic operators in the flavor multiplets as a result of the anomaly in the conservation of the supercurrent. These are specialized in section \ref{rigid} to rigid supersymmetry transformations on new minimal supergravity backgrounds that admit Killing spinors and the implications for supersymmetric observables are discussed. We conclude with a number of open questions in section \ref{discussion}. Appendix \ref{CS-review} contains a summary of the results of \cite{Papadimitriou:2019gel} for $\cn=1$ conformal supergravity, while in appendix \ref{WZ-proof} we provide the details of the Wess-Zumino consistency conditions calculation for the anomaly cocycles in the presence of flavor multiplets. Our spinor conventions follow those of \cite{Freedman:2012zz} and several useful gamma matrix identities can be found in appendix A of \cite{Papadimitriou:2019gel}.

\section{The local symmetry algebra of new minimal supergravity}
\label{nm-algebra}

We begin by reviewing some basic aspects of new minimal supergravity \cite{Akulov:1976ck,Sohnius:1981tp,Sohnius:1982fw,Ferrara:1988qxa}, including its local symmetry transformations and the corresponding algebra. As we will see, the gravity multiplet of new minimal supergravity can be formulated in terms of an effective gravity multiplet of $\cn=1$ conformal supergravity, allowing one to read off both the local symmetry algebra and the gravity multiplet anomalies directly from those of conformal supergravity computed in \cite{Papadimitriou:2019gel}.         

The field content of new minimal supergravity consists of the vielbein $e^a_\m$, an Abelian gauge field $A_\m$, an Abelian 2-form field $B_{\m\n}$ and a Majorana gravitino $\j_\m$, comprising $6+6$ bosonic and 12 fermionic off-shell degrees of freedom. Several properties of new minimal supergravity simplify when expressed in terms of the composite gauge field 
\be\label{effective-C}
C_\m\equiv A_\m-\frac32 V_\m,\qquad V_\m\equiv\frac14\e_\m{}^{\n\r\s}\Big(\pa_\n B_{\r\s}-\frac12\lbar\j_\n\g_\r\j_\s\Big).
\ee
Crucially, the composite field $C_\m$ transforms as a gauge field of $\cn=1$ conformal supergravity.

\subsection{Local symmetry transformations}

The local symmetries of new minimal supergravity are diffeomorphisms $\x^\m(x)$, local frame rotations $\l^{ab}(x)$, 0-form gauge transformations $\th(x)$, 1-form gauge transformations $\L_\m(x)$, and $\cq$-supersymmetry transformations $\ve(x)$. Under these the supergravity fields transform as\footnote{An interesting possibility is to promote the Abelian 0-form and 1-form symmetries of new minimal supergravity to a 2-group symmetry by modifying the gauge transformation of the 2-form field to include a term of the form \cite{Kapustin:2013uxa,Cordova:2018cvg,Benini:2018reh} 
\begin{equation*}
\d_\th B_{\m\n}=\frac{\k}{2\p}\th F_{\m\n},
\end{equation*}
where $F_{\m\n}=\pa_\m A_\n-\pa_\n A_\m$ and $\k$ is a constant. It would be interesting to determine whether the algebra can be adjusted to close off-shell in the presence of this deformation, and if so how the quantum anomalies would be modified. However, we will not consider this possibility in the present work.
}
\bal\label{NM-sugra-trans}
\d e^a_\m=&\;\x^\l\pa_\l e^a_\m+e^a_\l\pa_\m\x^\l-\l^a{}_b e^b_\m-\frac12\lbar\j_\m\g^a\ve,\NO\\
\d\j_\m=&\;\x^\l\pa_\l\j_\m+\j_\l\pa_\m\x^\l-\frac14\l_{ab}\g^{ab}\j_\m+\cd_\m\ve+\frac{i}{2}V^\n\g_\m\g_\n\g^5\ve-i\g^5\th\j_\m,\NO\\
\d A_\m=&\;\x^\l\pa_\l A_\m+A_\l\pa_\m\x^\l+\frac{i}{4}\lbar\ve\g_\m\g^{\r\s}\g^5\Big(\cd_\r\j_\s+\frac{i}{2}V^\n\g_\r\g_\n\g^5\j_\s\Big)+\pa_\m\th,\NO\\
\d B_{\m\n}=&\;\x^\l\pa_\l B_{\m\n}+B_{\l\n}\pa_\m\x^\l+B_{\m\l}\pa_\n\x^\l+\lbar\j_{[\m}\g_{\n]}\ve+\pa_\m\L_\n-\pa_\n\L_\m,
\eal
where the covariant derivatives of the gravitino and the spinor parameter $\ve$ are as in $\cn=1$ conformal supergravity and are given respectively in \eqref{covD-psi-phi} and \eqref{covD-parameters}. In new minimal supergravity, however, the gauge field $C_\m$ in the covariant derivatives is identified with the composite field \eqref{effective-C}.

Comparing the transformations \eqref{NM-sugra-trans} with those in $\cn=1$ conformal supergravity given in eq.~\eqref{sugra-trans} in appendix \ref{CS-review}, one notices that the transformation of the vielbein is the same in new minimal and conformal supergravity provided the Weyl transformation parameter $\s$ of conformal supergravity is set to zero. Similarly, the gravitino transformations coincide provided the Weyl parameter $\s$ and the $\cs$-supersymmetry parameter $\h$ of conformal supergravity are set to 
\be\label{CtoNM}
\s=0,\qquad \h=-\frac{i}{2}V^\r\g_\r\g^5\ve.
\ee
Using the following transformation of the composite vector field $V_\m$ defined in \eqref{effective-C}   
\be
\d V_\m=\x^\l\pa_\l V_\m+V_\l\pa_\m\x^\l-\frac14\e_{\m\n}{}^{\r\s}\lbar\ve\g^\n\cd_\r\j_\s+\frac14V^\n\lbar\ve\g_\m{}^\s\g_\n\j_\s+\frac12V^\n\lbar\ve\g_\n\j_\m,
\ee
the values \eqref{CtoNM} of the conformal supergravity parameters ensure also that the transformation of the composite gauge field $C_\m$ defined in \eqref{effective-C} coincides with that of the gauge field in conformal supergravity given in \eqref{sugra-trans}, namely  
\be
\d C_\m=\x^\l\pa_\l C_\m+C_\l\pa_\m\x^\l+\frac{3i}{4}\lbar\f_\m\g^5\ve-\frac{3i}{4}\lbar\j_\m\g^5\h+\pa_\m\th,
\ee
where $\f_\m$ is defined in \eqref{phi}. In summary, the fields $e^a_\m$, $\j_\m$ and $C_\m$ in new minimal supergravity transform exactly as the corresponding fields in $\cn=1$ conformal supergravity, provided the Weyl and $\cs$-supersymmetry parameters of conformal supergravity are set to the values in \eqref{CtoNM}. This observation allows us to deduce the local symmetry algebra of new minimal supergravity from the algebra of $\cn=1$ conformal supergravity.  

\subsection{Local symmetry algebra}

The relation between new minimal and conformal supergravities discussed above can be formulated as a map between the so called Ward operators of new minimal supergravity, $\d^{\rm NM}$, that generate the local symmetry transformations \eqref{NM-sugra-trans}, and those of conformal supergravity, $\d^C$. We have shown that the Ward operators of diffeomorphisms, local Lorentz and U(1) gauge transformations coincide in new minimal and conformal supergravities, namely 
\be
\d^{\rm NM}_\x=\d^{\rm C}_\x=\d_\x,\qquad \d^{\rm NM}_\l=\d^{\rm C}_\l=\d_\l,\qquad \d^{\rm NM}_\th=\d^{\rm C}_\th=\d_\th.
\ee
Moreover, the Ward operator of Weyl transformations is identically zero in new minimal supergravity, while the Ward operator of $\cq$-supersymmetry in new minimal supergravity is the sum of the $\cq$- and $\cs$-supersymmetry Ward operators in conformal supergravity, i.e.  
\be\label{Ward-map}
\d^{\rm NM}_\s=0,\qquad \d^{\rm NM}_\ve=\d^{\rm C}_\ve+\d^{\rm C}_{\h(\ve)},
\ee
with $\h(\ve)$ given in \eqref{CtoNM}. In addition, new minimal supergravity contains the Ward operator of 1-form gauge transformations $\d^{\rm NM}_\L$. It follows that all new minimal supergravity commutators that do not involve $\d^{\rm NM}_\L$ can be determined directly from the algebra of conformal supergravity, up to terms involving $\d^{\rm NM}_\L$. 

Let us first consider the commutator $[\d^{\rm NM}_\ve,\d^{\rm NM}_{\ve'}]$. Up to a possible contribution of $\d^{\rm NM}_\L$ on the r.h.s. that can be determined separately, this commutator can be read off from the algebra of conformal supergravity. Using \eqref{Ward-map} and the conformal supergravity algebra in \eqref{CS-local-algebra} we obtain
\be
[\d^{\rm NM}_\ve,\d^{\rm NM}_{\ve'}]=[\d^{\rm C}_\ve+\d^{\rm C}_\h,\d^{\rm C}_{\ve'}+\d^{\rm C}_{\h'}]=[\d^{\rm C}_\ve,\d^{\rm C}_{\ve'}]+[\d^{\rm C}_\ve,\d^{\rm C}_{\h'}]+[\d^{\rm C}_\h,\d^{\rm C}_{\ve'}]
=\d_\x+\d_{\l}+\d_{\th}+\d^{\rm C}_{\s},
\ee
where the field dependent parameters of the bosonic transformations on the r.h.s are given by 
\bal\label{NM-QQ-parameters}
\x^\m=&\;\frac12\lbar\ve'\g^\m\ve,\NO\\
\s=&\;\frac12(\lbar\ve\h'-\lbar\ve'\h)=-\frac{i}{4}V^\n\big(\lbar\ve\g_\n\g^5\ve'-\lbar\ve'\g_\n\g^5\ve\big)=0,\NO\\
\th=&\;-\frac12(\lbar\ve'\g^\n\ve)C_\n-\frac{3i}{4}\lbar\ve\g^5\h'+\frac{3i}{4}\lbar\ve'\g^5\h
=-\frac12(\lbar\ve'\g^\n\ve)A_\n,\NO\\
\l^a{}_b=&\;-\frac12(\lbar\ve'\g^\n\ve)\;\o_\n{}^a{}_b-\frac12\lbar\ve\g^a{}_b\h'+\frac12\lbar\ve'\g^a{}_b\h
=-\frac12(\lbar\ve'\g^\m\ve)\;\big(\o_\m{}^a{}_b+\e_{\m\n}{}^a{}_{b}V^\n\big).
\eal
Notice that the Weyl parameter $\s$ vanishes as required by the conditions \eqref{CtoNM}. In order to detect the possible presence of $\d^{\rm NM}_\L$ on the r.h.s. of the commutator $[\d^{\rm NM}_\ve,\d^{\rm NM}_{\ve'}]$ we need to evaluate it on $B_{\m\n}$. A straightforward calculation determines that 
\be
[\d^{\rm NM}_\ve,\d^{\rm NM}_{\ve'}]B_{\m\n}=(\d_\x+\d^{\rm NM}_\L)B_{\m\n},
\ee
where $\x^\m$ is as in \eqref{NM-QQ-parameters} and
\be
\L_\m=-\x_\m+B_{\m\n}\x^\n.
\ee

All remaining commutators either follow trivially from the corresponding ones in conformal supergravity, or they can be easily evaluated directly. Putting everything together, one finds that the non-vanishing commutators in new minimal supergravity are \cite{Sohnius:1981tp}\footnote{\label{effective-susy}The commutator $[\d^{\rm NM}_\ve,\d^{\rm NM}_{\ve'}]$ produces also a supersymmetry transformation with parameter $\ve''\sim\x^\m\j_\m$ \cite{Sohnius:1981tp,Sohnius:1982fw}. This term has no effect when working to leading order in the gravitino and so we do not include it in our analysis.}
\begin{alignat}{3}\label{NM-local-algebra}
&\;[\d_\x,\d_{\x'}]=\d_{\x''}, &&\x''^\m=\x^\n\pa_\n\x'^\m-\x'^\n\pa_\n\x^\m,\NO\\
&\;[\d_\l,\d_{\l'}]=\d_{\l''}, &&\l''^a{}_b=\l'^a{}_c\l^c{}_b-\l^a{}_c\l'^c{}_b,\rule{.0cm}{.5cm}\NO\\
&\;[\d^{\rm NM}_\ve,\d^{\rm NM}_{\ve'}]=\d_\x+\d_\l+\d_\th+\d^{\rm NM}_\L, \qquad &&\x^\m=\frac12\lbar\ve'\g^\m\ve,\quad \l^a{}_b=-\x^\m\big(\o_\m{}^a{}_b+\e_{\m\n}{}^a{}_{b}V^\n\big),\NO\\
& &&\;\th=-\x^\m A_\m,\quad \L_\m=-\x_\m+B_{\m\n}\x^\n.
\end{alignat}
The local parameters $\x^\m$, $\l^a{}_b$, $\th$ and $\ve$ transform as those in conformal supergravity with $\s=0$ (see eq.~\eqref{param-trans}), while the 1-form gauge parameter $\L_\m$ transforms as 
\be
\d\L_\m=\x^\n\pa_\n \L_\m+\L_\n\pa_\m\x^\n.
\ee
The algebra \eqref{NM-local-algebra} is the starting point for computing the candidate anomalies of new minimal supergravity by solving the corresponding Wess-Zumino consistency conditions.

\section{Ward identities and anomalies for the $R$-multiplet}
\label{nm-anomalies}

Supersymmetric theories with a U(1)$_R$ symmetry admit an $R$-multiplet \cite{Gates:1983nr}, which couples to new minimal background supergravity \cite{Komargodski:2010rb}. In this section we derive the Ward identities for the $R$-multiplet and we determine the corresponding bosonic and fermionic anomaly candidates. The relation between the local algebra of new minimal and conformal supergravity we identified in the previous section allows us to simply read off the $R$-multiplet anomalies from those of $\cn=1$ conformal supergravity found in \cite{Papadimitriou:2019gel}, without having to solve the Wess-Zumino consistency conditions for new minimal supergravity. 

\subsection{$R$-multiplet anomalies}

In four dimensions there are no genuine gravitational or Lorentz anomalies \cite{AlvarezGaume:1983ig}, and 1-form symmetries are also non anomalous.\footnote{A candidate 1-form symmetry anomaly of the form $\int e\;\L_\m\e^{\m\n\r\s}\pa_\n B_{\r\s}$ can be canceled by the local counterterm $\int B\wedge B$. See \cite{Brandt:1993vd,Brandt:1996au} for a classification of candidate anomalies in new minimal supergravity and \cite{Cordova:2018cvg} for a discussion of 1-form symmetry anomalies in connection to 2-group symmetries.} It follows that in a scheme (i.e. a choice of local counterterms) where the mixed axial-gravitational anomaly enters exclusively in the divergence of the $R$-current (see e.g. eq.~(2.43) of \cite{Jensen:2012kj}) the $R$-multiplet anomalies can be parameterized as 
\be\label{W-anomalies-NM}
\d_{\O_{\rm NM}} \mathscr{W}=\int d^4x\;e\;\big(-\th\ca^{\rm NM}_{R}-\lbar\ve\ca^{\rm NM}_{Q}\big),
\ee
where $\mathscr{W}[e,A,B,\j]$ is the generating functional of connected correlation functions of the $R$-multiplet currents and $\O_{\rm NM}=(\x,\l,\th,\L,\ve)$ denotes the set of local transformation parameters of new minimal supergravity. 

In the previous section we saw that $R$-symmetry transformations in new minimal and conformal supergravity coincide, while $\cq$-supersymmetry transformations in new minimal supergravity correspond to the sum of a $\cq$-supersymmetry and an $\cs$-supersymmetry transformation of an effective $\cn=1$ conformal supergravity, with gauge field as in \eqref{effective-C} and effective $\cs$-supersymmetry parameter $\h(\ve)$ as in \eqref{CtoNM}. It follows that the new minimal supergravity anomalies $\ca^{\rm NM}_R$ and $\ca^{\rm NM}_Q$ can be obtained directly from the anomalies of $\cn=1$ conformal supergravity. Namely, from \eqref{anomalies} we determine that
\bal\label{NM-anomalies}
\ca^{\rm NM}_R=&\;\ca^{\rm C}_R=\k^{(1)}\wt G G+\k^{(2)}\cp,\NO\\
\ca^{\rm NM}_{Q}=&\;\ca^{\rm C}_{Q}-\frac{i}{2}V^\r\g^5\g_\r\ca^{\rm C}_{S}=\k^{(1)} \ca^{(1)}_Q+\k^{(2)} \ca^{(2)}_Q,
\eal
where $\k^{(1)}$ and $\k^{(2)}$ are undetermined constants that depend on the specific theory that is placed on a background of new minimal supergravity, $G_{\m\n}=\pa_\m C_\n-\pa_\n C_\m$ is the fieldstrength of the composite gauge field $C_\m$, and $\wt G G$ and the Pontryagin density $\cp$ are defined respectively in \eqref{gauge-curvatures} and \eqref{metric-curvatures}. Moreover, the fermionic anomalies $\ca^{(1)}_Q$ and $\ca^{(2)}_Q$ are obtained from the fermionic anomalies $\ca^{\rm C}_{Q}$ and $\ca^{\rm C}_{S}$ in conformal supergravity through the identification \eqref{NM-anomalies} and take the form \cite{Papadimitriou:2019gel}   
\bal\label{NM-fermionic-anomalies}
\ca^{(1)}_Q=&\;-3i\wt G^{\m\n}C_\m\g^5\Big(\f_\n -\frac{i}{2}V^\r\g_\r\g^5\j_{\n}\Big)\NO\\
&\;-\frac{9i}{4}V^\k\g^5\g_\k\Big[\wt G^{\m\n}\cd_\m\j_{\n}+\frac{i}{2} G^{\m\n}\big(\g_{\m}{}^{[\s}\d_{\n}^{\r]}-\d_{\m}^{[\s}\d_{\n}^{\r]}\big)\g^5\cd_\r\j_\s+\frac{9}{4}P_{\m\n}g^{\m[\n}\g^{\r\s]}\cd_\r\j_\s\Big]+\co(\j^3),\NO\\
\rule{.0cm}{.9cm}\ca^{(2)}_Q=&\;-4\nabla_\m\big(C_\r \wt R^{\r\s\m\n}\big)\g_{(\n}\j_{\s)}+G_{\m\n} \wt R^{\m\n\r\s} \g_\r\j_\s-\frac{i}{2}V^\k\g^5\g_\k\Big[10i G^{\m\n}\big(\g_{\m}{}^{[\s}\d_{\n}^{\r]}-\d_{\m}^{[\s}\d_{\n}^{\r]}\big)\g^5\cd_\r\j_\s\NO\\
&\;+9 P_{\m\n}g^{\m[\n}\g^{\r\s]}\cd_\r\j_\s-3\Big(R^{\m\n\r\s}\g_{\m\n}-\frac12Rg_{\m\n}g^{\m[\n}\g^{\r\s]}\Big)\cd_\r\j_\s\Big]+\co(\j^3),
\eal
where the Schouten tensor $P_{\m\n}$ is defined in \eqref{Schouten}. At a fixed point the anomaly coefficients $\k^{(1)}$ and $\k^{(2)}$ are related to the $a$ and $c$ central charges as 
\be
\k^{(1)}=\frac{(5a-3c)}{27\p^2},\qquad \k^{(2)}=\frac{(c-a)}{24\p^2}.
\ee
The relation between the new minimal and conformal supergravity algebras we highlighted above ensures that the anomalies \eqref{NM-anomalies} are the general solution of the Wess-Zumino consistency conditions for the gravity multiplet of new minimal supergravity. 

It is possible that the supersymmetry anomalies \eqref{NM-fermionic-anomalies} are related to the superspace anomalies obtained in \cite{Bonora:1984pn,Bonora:2013rta} and \cite{Brandt:1993vd,Brandt:1996au} (see Type II anomalies in Table 9.1 of \cite{Brandt:1996au}). However, candidate anomalies in superspace and in components can differ because the extra auxiliary fields in the superspace formulation act as symmetry compensators \cite{Shamir:1992ff}. A known example of this phenomenon occurs in supersymmetric Yang-Mills theory in the presence of gauge anomalies. The superspace formulation of the theory does not exhibit a supersymmetry anomaly, but the component formulation in the Wess-Zumino gauge has a supersymmetry anomaly \cite{Itoyama:1985qi} (see also \cite{Piguet:1984aa,Guadagnini:1985ea, Zumino:1985vr}). This can be understood from the fact that in order to preserve the Wess-Zumino gauge, supersymmetry transformations require a compensating gauge transformation. If the theory has a gauge anomaly, this leads to a supersymmetry anomaly. However, gauge anomalies must be canceled for the consistency of supersymmetric Yang-Mills theory at the quantum level and so this fact has no physical significance. However, global symmetries such as $R$-symmetry or flavor symmetries can be anomalous and the supersymmetry anomaly they lead to is physical.

\subsection{Ward identities}

The Ward identities of the $R$-multiplet follow from the local symmetry transformations of new minimal supergravity \eqref{NM-sugra-trans} and the anomalous transformation \eqref{W-anomalies-NM} of the generating function. The form of the Ward identities is therefore independent of the specific quantum theory that is placed on a background of new minimal supergravity. All information about the microscopic theory is contained in the values of the anomaly coefficients $\k^{(1)}$ and $\k^{(2)}$.   

The fields of new minimal supergravity act as sources for the $R$-multiplet current operators, which are defined through a general variation of the generating function of connected correlators
\be\label{NM-variation-1}
\d\mathscr{W}=\int d^4x\;e\big(\d e^a_\m \<\ct^\m_a\>_s+\d A_\m\<\cj^\m\>_s+\d B_{\m\n}\<\ck^{\m\n}\>_s+\d\lbar\j_\m \<\cs^\m\>_s\big),
\ee
so that
\be\label{NM-currents-1}
\<\ct^\m_a\>_s=e^{-1}\frac{\d\mathscr{W}}{\d e^a_\m},
\qquad \<\cj^\m\>_s=e^{-1}\frac{\d\mathscr{W}}{\d A_\m},\qquad \<\ck^{\m\n}\>_s=e^{-1}\frac{\d\mathscr{W}}{\d B_{\m\n}},\qquad
\<\cs^\m\>_s=e^{-1}\frac{\d\mathscr{W}}{\d\lbar\j_\m},
\ee
where $e\equiv \det(e^a_\m)$ and $\<\cdots\>_s$ denotes a (connected) correlation function in the presence of arbitrary sources. In particular, any $n$-point function involving $R$-multiplet currents can be obtained by further differentiating these expressions with respect to the corresponding sources. 

A slightly different set of $R$-multiplet operators is often defined by parameterizing a general variation of the generating functional as \cite{Sohnius:1981tp} (see also \cite{Komargodski:2010rb})
\be\label{NM-variation-2}
\d\mathscr{W}=\int d^4x\;e\big(\d e^a_\m \<\Hat\ct^\m_a\>_s+\d C_\m\<\Hat\cj^\m\>_s+\d B_{\m\n}\<\Hat\ck^{\m\n}\>_s+\d\lbar\j_\m \<\Hat\cs^\m\>_s\big),
\ee
so that the $R$-current couples to the composite gauge field $C_\m$ rather than to $A_\m$. The two sets of operators are related through spectral flow:
\bal\label{spectral-flow}
\<\Hat\ct^\m_a\>_s=&\;\<\ct^\m_a\>_s+\frac32\Big(V_ag^{\m\n}+V^\m e^\n_a -V^\n e^\m_a -\frac18\e^{\m\n\r\s}\lbar\j_\r\g_a\j_\s\Big)\<\cj_\n\>_s,\NO\\
\<\Hat\ck^{\m\n}\>_s=&\;\<\ck^{\m\n}\>_s+\frac38\e^{\m\n\r\s}\pa_\r\<\cj_\s\>_s,\NO\\
\<\Hat\cs^\m\>_s=&\;\<\cs^\m\>_s+\frac38\e^{\m\n\r\s}\g_\r\j_\s\<\cj_\n\>_s,\NO\\
\<\Hat\cj^\m\>_s=&\;\<\cj^\m\>_s.
\eal
Besides obeying simpler Ward identities, the advantage of the hatted operators is that they couple also to conformal supergravity and are therefore appropriate for describing superconformal theories.  

In order to derive the Ward identities of the $R$-multiplet we equate the anomalous transformation \eqref{W-anomalies-NM} of the generating function with either \eqref{NM-variation-1} or \eqref{NM-variation-2}, evaluated on the symmetry transformations \eqref{NM-sugra-trans} of new minimal supergravity. In terms of the hatted currents the resulting Ward identities take the form
\bal
\label{NM-WardIDs}
&e^a_\m\nabla_\n\<\Hat\ct^\n_a\>_s+\nabla_\n(\lbar\j_\m  \<\Hat\cs^\n\>_s)-\lbar\j_\n\overleftarrow \cd_\m \<\Hat\cs^\n\>_s-G_{\m\n}\<\Hat\cj^\n\>_s-H_{\m\r\s}\<\Hat\ck^{\r\s}\>_s\NO\\
&\hspace{1.cm}+2B_{\m\s}\de_\r\<\Hat\ck^{\r\s}\>_s+C_\m\big(\nabla_\n\<\Hat\cj^\n\>_s+i\lbar\j_\n \g^5\<\Hat\cs^\n\>_s\big)-\o_\m{}^{ab}\Big(e_{\n [a}\<\Hat\ct^\n_{b]}\>_s+\frac14\lbar\j_\n\g_{ab}\<\Hat\cs^\n\>_s\Big)=0,\NO\\
\rule{.0cm}{.7cm}&e_{\m[a} \<\Hat\ct^\m_{b]}\>_s+\frac14\lbar\j_\m\g_{ab} \<\Hat\cs^\m\>_s=0,\NO\\
\rule{.0cm}{.7cm}&\nabla_\m\<\Hat\ck^{\m\n}\>_s=0,\NO\\
\rule{.0cm}{.7cm}&\nabla_\m \<\Hat\cj^\m\>_s+i\lbar\j_\m\g^5 \<\Hat\cs^\m\>_s=\ca_R^{\rm NM},\\
\rule{.0cm}{.7cm}&\Big(\cd_\m -\frac{i}{2}V^\r\g^5\g_\r\g_\m \Big)\<\Hat\cs^\m\>_s-\frac12\g^a\j_\m \<\Hat\ct^\m_a\>_s-\frac{3i}{4}\g^5\Big(\f_\m-\frac{i}{2}V^\r\g_\r\g^5\j_\m\Big)\<\Hat\cj^\m\>_s-\g_{[\m}\j_{\n]}\<\Hat\ck^{\m\n}\>_s=\ca_Q^{\rm NM},\NO
\eal
where the fieldstrength $H_{\m\n\r}$ of the 2-form gauge field $B_{\m\n}$ is given by
\be
H_{\m\n\r}=\pa_\m B_{\n\r}+\pa_\r B_{\m\n}+\pa_\n B_{\r\m}.
\ee

The Ward identities are slightly more cumbersome in terms of the currents \eqref{NM-currents-1}, namely 
\bal
\label{NM-WardIDs-unhatted}
&e^a_\m\nabla_\n\<\ct^\n_a\>_s+\nabla_\n(\lbar\j_\m  \<\cs^\n\>_s)-\lbar\j_\n\overleftarrow \cd_\m \<\cs^\n\>_s-\frac{3i}{2}V_\m\lbar\j_\n \g^5\<\cs^\n\>_s-F_{\m\n}\<\cj^\n\>_s-H_{\m\r\s}\<\ck^{\r\s}\>_s\NO\\
&\hspace{1.cm}+2B_{\m\s}\de_\r\<\ck^{\r\s}\>_s+A_\m\big(\nabla_\n\<\cj^\n\>_s+i\lbar\j_\n \g^5\<\cs^\n\>_s\big)-\o_\m{}^{ab}\Big(e_{\n [a}\<\ct^\n_{b]}\>_s+\frac14\lbar\j_\n\g_{ab}\<\cs^\n\>_s\Big)=0,\NO\\
\rule{.0cm}{.7cm}&e_{\m[a} \<\ct^\m_{b]}\>_s+\frac14\lbar\j_\m\g_{ab} \<\cs^\m\>_s=0,\NO\\
\rule{.0cm}{.7cm}&\nabla_\m\<\ck^{\m\n}\>_s=0,\NO\\
\rule{.0cm}{.7cm}&\nabla_\m \<\cj^\m\>_s+i\lbar\j_\m\g^5 \<\cs^\m\>_s=\ca_R^{\rm NM},\NO\\
\rule{.0cm}{.7cm}&\Big(\cd_\m -\frac{i}{2}V^\r\g^5\g_\r\g_\m \Big)\<\cs^\m\>_s-\frac12\g^a\j_\m \<\ct^\m_a\>_s+\frac{i}{4}\g_\m\g^{\r\s}\g^5\Big(\cd_\r\j_\s+\frac{i}{2}V^\n\g_\r\g_\n\g^5\j_\s\Big)\<\cj^\m\>_s\NO\\
&\hspace{1.cm}-\g_{[\m}\j_{\n]}\<\ck^{\m\n}\>=\ca_Q^{\rm NM}.
\eal
These can be deduced by inserting the expressions \eqref{spectral-flow} for the hatted currents in the Ward identities \eqref{NM-WardIDs}, but it is technically significantly simpler to obtain them directly from the variation \eqref{NM-variation-1} of the generating function. 

We emphasize that the Ward identities \eqref{NM-WardIDs} or \eqref{NM-WardIDs-unhatted} involve one-point functions in the presence of arbitrary sources, i.e. generic background fields. This means that differentiating these identities with respect to the background fields and using the definitions of the current operators above one can derive the Ward identities for any correlation function of $R$-multiplet currents, both in flat space and on any new minimal supergravity background. In particular, the anomalies $\ca_R^{\rm NM}$ and $\ca_Q^{\rm NM}$ contribute contact terms in certain flat space higher-point functions \cite{Katsianis:2019hhg,followup}.

\section{Ward identities and anomalies in the presence of flavor symmetries}
\label{nm-vector-multiplet}

Supersymmetric field theories may possess additional global symmetries beyond those encoded in the gravity multiplet. In order to derive the Ward identities and their quantum anomalies in the presence of such flavor symmetries we need to couple the gravity multiplet of new minimal supergravity to a number of vector multiplets (gauge multiplets in the terminology of \cite{Sohnius:1982fw}). In this section we will consider an arbitrary number $N$ of Abelian vector multiplets $(a^I_\m, \l^I, D^I)$, $I=1,\ldots, N$. The subsequent analysis can be easily generalized to non Abelian vector multiplets, but we will not address this case here.

The local symmetry transformations of the vector multiplet fields take the from \cite{Sohnius:1982fw}
\bal\label{NM-vector-trans}
\d a^I_\m=&\;\x^\n\pa_\n a^I_\m+a^I_\n\pa_\m\x^\n+\frac12\lbar\ve\g_\m\l^I+\pa_\m\o^I,\NO\\
\d\l^I=&\;\x^\n\pa_\n\l^I-\frac14\l_{ab}\g^{ab}\l^I-\frac14\big(\g^{\r\s}\mathscr{F}^I_{\r\s}+\g^5D^I\big)\ve-i\th\g^5\l^I,\NO\\
\d D^I=&\;\x^\n\pa_\n D^I+\lbar\ve\g^5\g^\m\Big[\cd_\m\l^I+\frac14\big(\g^{\r\s}\mathscr{F}^I_{\r\s}+\g^5D^I\big)\j_\m\Big],
\eal
where $f_{\m\n}^I=\pa_\m a_\n^I-\pa_\n a_\m^I$ is the flavor fieldstrength with 
\be
\mathscr{F}^I_{\m\n}=f^I_{\m\n}-\lbar\j_{[\m}\g_{\n]}\l^I,
\ee
and the covariant derivative acts on the flavorinos as
\be
\cd_\m\l^I=\Big(\pa_\m+\frac14\o_{\m ab}\g^{ab}+i\g^5C_\m\Big)\l^I.
\ee

A straightforward but tedious calculation shows that these transformations form another off-shell representation of the new minimal supergravity algebra \eqref{NM-local-algebra}, except that the commutator between two supersymmetry transformations has an additional term, namely 
\be\label{susysusy-extended}
[\d_\ve,\d_{\ve'}]=\d_\x+\d_\l+\d_\th+\d_\L+\d_{\o},
\ee
where the composite parameters $\x^\m$, $\l^a{}_b$, $\th$ and $\L_\m$ are as in \eqref{NM-local-algebra}, while the flavor transformation parameter takes the form  
\be
\o^I=-\x^\m a^I_\m.
\ee
As before, we are neglecting a supersymmetry transformation on the r.h.s. of \eqref{susysusy-extended} that plays no role to leading order in the fermions (see footnote \ref{effective-susy}).

\subsection{$R$-multiplet anomalies with flavors}
\label{flavor-anomalies}

In the presence of flavors, the anomalous transformation of the generating functional of connected correlators $\mathscr{W}[e,A,B,\j,a^I,\l^I,D^I]$ under the extended local symmetries $\O=(\x,\l,\th,\L,\ve,\o^I)$ can be parameterized as
\be\label{W-anomalies-NM-vector}
\d_{\O} \mathscr{W}=\int d^4x\;e\;\big(-\th\ca_{R}-\o^I\ca_I-\lbar\ve\ca_{Q}\big),
\ee
where the $R$-symmetry and supersymmetry anomalies now receive additional contributions relative to the gravity multiplet anomalies \eqref{NM-anomalies} due to the flavors, and there is a new anomaly in the flavor gauge transformations. 

Turning on background fields for the flavor multiplets leads to several independent solutions of the Wess-Zumino consistency conditions, in addition to the two gravity multiplet cocycles $\k^{(1)}$ and $\k^{(2)}$. The $R$-symmetry and flavor anomalies take the form \cite{Brandt:1993vd,Brandt:1996au,Anselmi:1997am,Anselmi:1997ys,Intriligator:2003jj,Cassani:2013dba,Assel:2014tba}
\bal\label{NM-anomalies-vector-bosonic}
\ca_R=&\;\k^{(1)}\wt G G+\k^{(2)}\cp+\a^{(4)}_I\wt Ff^I+\big(\k^{(5)}_{(IJ)}-\a^{(5)}_{(IJ)}\big)\wt f^I f^J+\k^{(7)}_I\big(D^I-i\e^{\m\n\r\s}a^I_\m \pa_\n B_{\r\s}+\lbar\l^I\g^5\g^\m\j_\m\big),\NO\\
\rule{.0cm}{.7cm}\ca_I=&\;\k^{(3)}_I\Big(\cp-\frac83G\wt G\Big)+\big(\k^{(4)}_I-\a^{(4)}_I\big)\wt FF+\a^{(5)}_{(IJ)}\wt F f^J+\k^{(6)}_{(IJK)}\wt f^J f^K\NO\\
&+\k^{(7)}_I\Big[i\e^{\m\n\r\s}A_\m \pa_\n B_{\r\s}-\frac{i}{2}R-3iV_\m V^\m-\frac{i}{2}\de_\n(\lbar\j^\n\g^\m\j_\m)+\frac{i}{2}\lbar\j_\m\g^{\m\r\s}\Big(\cd_\r\j_\s+\frac{3i}{4}V^\t\g_\r\g_\t\g^5\j_\s\Big)\Big]\NO\\
&+\k^{(8)}_{[IJ]}\big(D^J-i\e^{\m\n\r\s}a^J_\m \pa_\n B_{\r\s}+\lbar\l^J\g^5\g^\m\j_\m\big),
\eal
where the notation for $f^I_{\m\n}$ and $F_{\m\n}$ is analogous to that for $G_{\m\n}$ in \eqref{gauge-curvatures} and summation over repeated flavor indices is implicit. Besides the anomaly coefficients $\k^{(1)}$ and $\k^{(2)}$ of the gravity multiplet, there are six additional anomaly coefficients in the presence flavors that cannot be eliminated by local counterterms. The goal of this section is to determine the supersymmetry anomaly $\ca_Q$ corresponding to all flavor anomaly coefficients in \eqref{NM-anomalies-vector-bosonic}.

Before we turn to the supersymmetry anomaly, several comments are in order regarding the structure of the flavor anomalies in \eqref{NM-anomalies-vector-bosonic}. Firstly, the flavor 't Hooft anomaly coefficients can be expressed in terms of the $R$-charges ${\bf R}$ of the microscopic theory fermions and their charges ${\bf F}_I$ under the flavor symmetries. In particular, the first flavor coefficient takes the form  $\k_I^{(3)}\sim\Tr{\bf F}_I$, while $\k_I^{(4)}\sim\Tr({\bf R}^2 {\bf F}_I)$ is only independent for massive theories, since at a superconformal fixed point $\k_I^{(4)}\sim \k_I^{(3)}$ -- see eq.~(1.5) in \cite{Intriligator:2003jj}. Secondly, the $\k_I^{(4)}$ cocycle can alternatively be expressed as   
\be
\left.\ca_R\right|_{\k^{(4)}}=\frac32\k_{I}^{(4)}\e^{\m\n\r\s}\pa_\m V_\n f^I_{\r\s},\qquad \left.\ca_I\right|_{\k^{(4)}}=\k_{I}^{(4)}\wt G G,
\ee
by means of a local counterterm. Hence, the coefficients of the Pontryagin density, $\cp$, and of $G\wt G$ in $\ca_I$ are independent for non  conformal theories.\footnote{I thank Cyril Closset for pointing this out to me.} The anomaly coefficients $\k_{(IJ)}^{(5)}$ and $\k_{(IJK)}^{(6)}$ are totally symmetric in the flavor indices and are proportional to $\Tr({\bf R}{\bf F}_{(I}{\bf F}_{J)})$ and $\Tr({\bf F}_{(I}{\bf F}_J{\bf F}_{K)})$, respectively. These cocycles often appear in the literature together with a term bilinear in the flavorinos (gauginos) $\l^I$ -- see e.g. eq.~(20.71) in \cite{West:1990tg}. Such expressions differ from the ones given in \eqref{NM-anomalies-vector-bosonic} by local counterterms of the form $\k^{(5)}_{(IJ)}\int d^4x\;e\;A_\m\lbar\l^I\g^\m\l^J$ and $\k^{(6)}_{(IJK)}\int d^4x\;e\; a^I_\m\lbar\l^J\g^\m\l^K$, respectively. Another set of local counterterms that is useful in order to compare with the expressions for the $\k^{(4)}$ and $\k^{(5)}$ cocycles in the literature is 
\be\label{scheme-ct}
\mathscr{W}_{ct}=-\a^{(4)}_I\int d^4x\;e\;\e^{\m\n\r\s}a_\m^IA_\n F_{\r\s}-\a^{(5)}_{(IJ)}\int d^4x\;e\;\e^{\m\n\r\s}A_\m a_\n^I f^J_{\r\s},
\ee
where $\a^{(4)}_I$ and $\a^{(5)}_{(IJ)}$ are arbitrary constants. These local counterterms can be used to move the corresponding anomalies between the divergence of the $R$-current and the divergence of the flavor currents and have been included in the expressions for the flavor anomalies in \eqref{NM-anomalies-vector-bosonic}. 

Finally, the Fayet-Iliopoulos type cocycles $\k^{(7)}$ and $\k^{(8)}$ were found in \cite{Brandt:1993vd} and their contribution to the supersymmetry anomaly was already given there (in the case of the $\k^{(7)}$ cocycle only implicitly). It would be interesting to explore the significance of these cocycles; we are unaware of any computation of these coefficients in specific theories. Notice that the coefficients $\k^{(8)}_{[IJ]}$ are antisymmetric in the flavor indices and so this cocycle can only exist in the presence of at least two flavors. Moreover, the total derivative term bilinear in the gravitino in the $\k^{(7)}$ cocycle can be removed by a local counterterm of the form $\k^{(7)}_I\int d^4 x\;e\;a_\n^I\lbar\j^\n\g^\m\j_\m$. However, this would modify the form of the supersymmetry anomaly $\ca_Q^{(7)}$ given in eq.~\eqref{NM-anomalies-fermionic-vector-components}.

The supersymmetry anomaly is determined by the Wess-Zumino consistency conditions \cite{Wess:1971yu}
\be\label{WZ-condition}
[\d_\O,\d_{\O'}] \mathscr{W}=\d_{[\O,\O']}\mathscr{W},
\ee
for any pair of local symmetries $\O=(\x,\l,\th,\L,\ve,\o^I)$ and $\O'=(\x',\l',\th',\L',\ve',\o'^I)$. Writing  
\bal\label{NM-anomalies-vector-fermionic}
\ca_{Q}=&\;\sum_{i=1}^8\k^{(i)}_{\{IJ\ldots\}} \ca^{(i)\{IJ\ldots\}}_Q+\ca_Q^{ct},
\eal
the Wess-Zumino consistency conditions can be solved independently for each cocycle, i.e. for each anomaly coefficient. In section \ref{nm-anomalies} we already determined the gravity multiplet supersymmetry anomalies $\ca^{(1)}_Q$ and $\ca^{(2)}_Q$ in eq.~\eqref{NM-fermionic-anomalies} by embedding the new minimal supergravity algebra in the algebra of $\cn=1$ conformal supergravity and utilizing the results of \cite{Papadimitriou:2019gel}. In appendix \ref{WZ-proof} we solve the Wess-Zumino consistency conditions for each of the six flavor cocycles using as input the bosonic anomalies \eqref{NM-anomalies-vector-bosonic}. The resulting fermionic anomalies take the form
\bal\label{NM-anomalies-fermionic-vector-components}
\ca_Q^{(3)I}=&\;8i\wt G^{\m\n}a^I_\m\g^5\Big(\f_\n -\frac{i}{2}V^\r\g_\r\g^5\j_{\n}\Big)-4\nabla_\m\big(a^I_\r \wt R^{\r\s\m\n}\big)\g_{(\n}\j_{\s)}\NO\\
&\;\hspace{2.cm}+f^I_{\m\n} \wt R^{\m\n\r\s} \g_\r\j_\s+4iV^\r\wt f^{I\m\n}\g^5\g_\r\cd_\m\j_{\n}+\co(\{\j,\l\}^3),\NO\\
\rule{.0cm}{.7cm}\ca_Q^{(4)I}=&\;-i\wt F^{\m\n}a_\m^I\g_\n\g^{\r\s}\g^5\Big(\cd_\r\j_\s+\frac{i}{2}V^\t\g_\r\g_\t\g^5\j_\s\Big)+\co(\{\j,\l\}^3),\NO\\
\rule{.0cm}{.7cm}\ca_Q^{(5)(IJ)}=&\;-2A^\m\wt f^{(I}_{\m\n}\g^\n\l^{J)}+\co(\{\j,\l\}^3),\NO\\
\rule{.0cm}{.7cm}\ca_Q^{(6)(IJK)}=&\;-2\wt f^{(I\m\n}a^J_\m\g_\n\l^{K)}+\co(\{\j,\l\}^3),\NO\\
\rule{.0cm}{.7cm}\ca_Q^{(7)I}=&\;-A_\m\big(\g^5\g^\m\l^{I}+i\e^{\m\n\r\s}a_\n^{I}\g_\r\j_\s\big)-\frac{i}{2} a_\m^I\g^\m\g^{\r\s}\Big(\cd_\r\j_\s+\frac{i}{2}V^\t\g_\r\g_\t\g^5\j_\s\Big)+\co(\{\j,\l\}^3),\NO\\
\rule{.0cm}{.7cm}\ca_Q^{(8)[IJ]}=&\;-a_\m^{[I}\Big(\g^5\g^\m\l^{J]}+\frac{i}{2}\e^{\m\n\r\s}a_\n^{J]}\g_\r\j_\s\Big)+\co(\{\j,\l\}^3),
\eal
where $\co(\{\j,\l\}^3)$ is shorthand for $\co(\j^3,\j^2\l,\j\l^2,\l^3)$. Moreover, the contribution of the local counterterms \eqref{scheme-ct} to the supersymmetry anomaly is
\bal\label{scheme-AQ}
\ca_Q^{ct}=&\;\a^{(4)}_I\Big[\frac{i}{2}\big(2\wt F^{\m\n}a_\m^I-\wt f^{I\m\n}A_\m\big)\g_\n\g^{\r\s}\g^5\Big(\cd_\r\j_\s+\frac{i}{2}V^\t\g_\r\g_\t\g^5\j_\s\Big)-\wt F^{\m\n}A_\m\g_\n\l^I\Big]\\
&\;+\a^{(5)}_{(IJ)}\Big[2A^\m\wt f^{(I}_{\m\n}\g^\n\l^{J)}-\wt F^{\m\n}a_\m^{(I}\g_\n\l^{J)}-\frac{i}{2}\wt f^{(I\m\n}a_\m^{J)}\g_\n\g^{\r\s}\g^5\Big(\cd_\r\j_\s+\frac{i}{2}V^\t\g_\r\g_\t\g^5\j_\s\Big)\Big].\NO
\eal

Some of these contributions to the supersymmetry anomaly have been discussed in the literature before. As we mentioned above, the supersymmetry anomalies $\ca_Q^{(7)}$ and $\ca_Q^{(8)}$ were obtained in \cite{Brandt:1993vd}. $\ca_Q^{(5)}$ was pointed out in \cite{Shamir:1992ff}, while $\ca_Q^{(6)}$ is the Abelian (and global) analogue of the supersymmetry anomaly in super Yang-Mills theory in the presence of a gauge anomaly discussed in \cite{Itoyama:1985qi} (see also \cite{Piguet:1984aa,Guadagnini:1985ea, Zumino:1985vr}).\footnote{An analogous supersymmetry anomaly was found in the presence of a gravitational anomaly in two-dimensional theories in \cite{Howe:1985uy,Itoyama:1985ni,Tanii:1985wy}.} We are not aware of any earlier work where $\ca_Q^{(3)}$ or $\ca_Q^{(4)}$ were obtained. Notice that the anomalies $\ca_Q^{(3)}$, $\ca_Q^{(6)}$ and $\ca_Q^{(8)}$ are related only to the flavor anomalies and imply that supersymmetry can be anomalous even if $R$-symmetry is not.

Except for the Fayet-Iliopoulos type anomalies $\k^{(7)}_I$ and $\k^{(8)}_{[IJ]}$, the non covariant part of the supersymmetry anomalies in \eqref{NM-anomalies-fermionic-vector-components} is directly related to the Chern-Simons forms of the corresponding $R$-symmetry and flavor anomalies \cite{Itoyama:1985qi}. Writing these in terms of Chern-Simons forms we have 
\be
\d_\th\mathscr{W}=-\int \th\;\tx d\mathscr{Q}^{\rm CS}=\int \tx d\th\wedge\mathscr{Q}^{\rm CS},
\ee
and similarly 
\be
\d_\o\mathscr{W}=-\int \o^I\;\tx d\mathscr{Q}_I^{\rm CS}=\int \tx d\o^I\wedge\mathscr{Q}_I^{\rm CS}.
\ee
From the Wess-Zumino consistency conditions $[\d_\th,\d_\ve]\mathscr{W}=0$ and $[\d_\o,\d_\ve]\mathscr{W}=0$ follows that 
\be
\d_\th\d_\ve\mathscr{W}=\d_\ve\d_\th\mathscr{W}=\int \tx d\th\wedge\d_\ve\mathscr{Q}^{\rm CS},\qquad \d_\o\d_\ve\mathscr{W}=\d_\ve\d_\o\mathscr{W}=\int \tx d\o^I\wedge\d_\ve\mathscr{Q}_I^{\rm CS}.
\ee
Hence,
\be
\d_\ve\mathscr{W}=\int (A\wedge\d_\ve\mathscr{Q}_{\rm CS}+ a^I\wedge\d_\ve\mathscr{Q}_I^{\rm CS}+\text{covariant})\equiv-\int d^4x\;e\;\lbar\ve\ca_Q,
\ee
where the covariant part of the supersymmetry anomaly is invariant under both $R$-symmetry and flavor gauge transformations. The Chern-Simons forms are not sufficient to characterize the covariant part of the supersymmetry anomaly, but it can be determined by the Wess-Zumino consistency condition $[\d_{\ve},\d_{\ve'}]\mathscr{W}=(\d_\th+\d_\o)\mathscr{W}$ with $\th=-\frac12(\lbar\ve'\g^\m\ve) A_\m$, $\o^I=-\frac12(\lbar\ve'\g^\m\ve) a^I_\m$.  From the analysis in appendix \ref{WZ-proof} we find that the covariant part of the supersymmetry anomalies $\ca_Q^{(4)I}$, $\ca_Q^{(5)(IJ)}$ and $\ca_Q^{(6)(IJK)}$ is cubic in the fermions, which is why only the non covariant part related to the Chern-Simons forms appears in the corresponding expressions in \eqref{NM-anomalies-fermionic-vector-components}. However, the covariant part of $\ca_Q^{(3)I}$, as well as of the gravity multiplet anomalies $\ca^{(1)}_Q$ and $\ca^{(2)}_Q$, contains terms linear in the fermions.

\subsection{Ward identities}

The vector multiplet fields act as sources for the local operators in the flavor multiplets:
\be\label{vector-operators}
\<j^\m_I\>_s=e^{-1}\frac{\d\mathscr{W}}{\d a^I_\m},\qquad \<\co_I\>_s=e^{-1}\frac{\d\mathscr{W}}{\d D^I},\qquad
\<\c_I\>_s=e^{-1}\frac{\d\mathscr{W}}{\d\lbar\l^I}.
\ee
Using these operators and the local symmetry transformations \eqref{NM-vector-trans} in the anomalous transformation of the generating functional in \eqref{W-anomalies-NM-vector} leads to the general Ward identities for the $R$-multiplet in the presence of flavor symmetries, generalizing \eqref{NM-WardIDs}:
\bal
\label{NM-WardIDs-flavors}
&e^a_\m\nabla_\n\<\Hat\ct^\n_a\>_s+\nabla_\n(\lbar\j_\m  \<\Hat\cs^\n\>_s)-\lbar\j_\n\overleftarrow \cd_\m \<\Hat\cs^\n\>_s-G_{\m\n}\<\Hat\cj^\n\>_s-H_{\m\r\s}\<\Hat\ck^{\r\s}\>_s\NO\\
&-\lbar\l^I\overleftarrow \cd_\m \<\c_I\>_s-f_{\m\n}^I\<j^\n_I\>_s-\pa_\m D^I\<\co_I\>_s+2B_{\m\s}\de_\r\<\Hat\ck^{\r\s}\>_s+a_\m^I\de_\n\<j^\n_I\>_s\NO\\
&\hspace{.cm}+C_\m\big(\nabla_\n\<\Hat\cj^\n\>_s+i\lbar\j_\n \g^5\<\Hat\cs^\n\>_s+i\lbar\l^I\g^5\<\c_I\>_s\big)-\o_\m{}^{ab}\Big(e_{\n [a}\<\Hat\ct^\n_{b]}\>_s+\frac14\lbar\j_\n\g_{ab}\<\Hat\cs^\n\>_s+\frac14\lbar\l^I\g_{ab} \<\c_I\>_s\Big)=0,\NO\\
\rule{.0cm}{.7cm}&e_{\m[a} \<\Hat\ct^\m_{b]}\>_s+\frac14\lbar\j_\m\g_{ab} \<\Hat\cs^\m\>_s+\frac14\lbar\l^I\g_{ab} \<\c_I\>_s=0,\NO\\
\rule{.0cm}{.7cm}&\nabla_\m\<\Hat\ck^{\m\n}\>_s=0,\NO\\
\rule{.0cm}{.7cm}&\nabla_\m \<\Hat\cj^\m\>_s+i\lbar\j_\m\g^5 \<\Hat\cs^\m\>_s+i\lbar\l^I\g^5\<\c_I\>_s=\ca_R,\NO\\
\rule{.0cm}{.6cm}&\nabla_\m \<j^\m_I\>_s=\ca_I,\NO\\
\rule{.0cm}{.7cm}&\Big(\cd_\m -\frac{i}{2}V^\r\g^5\g_\r\g_\m \Big)\<\Hat\cs^\m\>_s-\frac12\g^a\j_\m \<\Hat\ct^\m_a\>_s-\frac{3i}{4}\g^5\Big(\f_\m-\frac{i}{2}V^\r\g_\r\g^5\j_\m\Big)\<\Hat\cj^\m\>_s-\g_{[\m}\j_{\n]}\<\Hat\ck^{\m\n}\>_s\\
&-\frac12\g_\m\l^I\<j^\m_I\>_s+\frac14\big(-\g^{\r\s}\mathscr{F}_{\r\s}+\g^5D^I\big)\<\c_I\>_s-\g^5\g^\m\Big[\cd_\m\l^I+\frac14\big(\g^{\r\s}\mathscr{F}_{\r\s}+\g^5D^I\big)\j_\m\Big]\<\co_I\>_s=\ca_Q.\NO
\eal

\section{Anomalous supersymmetry transformations of the fermionic operators}
\label{supercurrent}

An important consequence of the supersymmetry anomaly \eqref{NM-anomalies-vector-fermionic} is that it leads to an anomalous supersymmetry transformation for the fermionic operators in the gravity and flavor multiplets \cite{Papadimitriou:2017kzw,An:2017ihs,An:2018roi,Papadimitriou:2019gel}. As we review in the next section, when restricted to a specific background admitting Killing spinors, the anomalous terms in the rigid supersymmetry transformation of the fermionic operators depend on the {\em bosonic} background and have physical implications. In particular, the anomalous transformation of the supercurrent leads to a deformed supersymmetry algebra.  

The transformations of the $R$-multiplet currents and of the flavor multiplet operators under the local symmetries of new minimal supergravity are directly related with the Ward identities \eqref{NM-WardIDs-flavors}. These correspond to first class constraints on the symplectic space of couplings and local operators, generating the local symmetry transformations under the Poisson bracket \cite{Papadimitriou:2016yit}. In particular, the quantum transformations of the operators are encoded in the anomalies of the Ward identities. This method was used in appendix B.1 of \cite{Papadimitriou:2017kzw} in order to obtain the anomalous transformation of the supercurrent under $\cq$- and $\cs$-supersymmetry in conformal supergravity for the case $a=c$.    

An alternative way to determine the transformation of the quantum operators under the local symmetries is to use their defining relation in terms of the generating function. For example, under local supersymmetry transformations the supercurrent and the fermionic operator in the flavor multiplets transform respectively as
\bal\label{operator-trans-general}
\d_\ve\<\Hat\cs^\m\>_s=&\;e^{-1}\d_\ve\Big(\frac{\d}{\d\lbar\j_\m}\Big)\mathscr{W}+e^{-1}\frac{\d}{\d\lbar\j_\m}\d_\ve\mathscr{W}=e^{-1}\d_\ve\Big(\frac{\d}{\d\lbar\j_\m}\Big)\mathscr{W}-e^{-1}\frac{\d}{\d\lbar\j_\m}\int d^4x\;e\;\lbar\ve\ca_Q,\NO\\
\rule{.0cm}{1.cm}\d_\ve\<\c_I\>_s=&\;e^{-1}\d_\ve\Big(\frac{\d}{\d\lbar\l^I}\Big)\mathscr{W}+e^{-1}\frac{\d}{\d\lbar\l^I}\d_\ve\mathscr{W}=e^{-1}\d_\ve\Big(\frac{\d}{\d\lbar\l^I}\Big)\mathscr{W}-e^{-1}\frac{\d}{\d\lbar\l^I}\int d^4x\;e\;\lbar\ve\ca_Q.
\eal
The transformation of the functional derivatives determines the classical transformation of the operators and follows directly from the classical symmetry transformations of new minimal supergravity. In particular, from \eqref{NM-sugra-trans} and \eqref{NM-vector-trans} we obtain 
\bal\label{operator-trans-cl}
\d_\ve\Big(\frac{\d}{\d\lbar\j_\m}\Big)
=&\;\frac{1}{2}\g^a\ve\frac{\d}{\d e^a_\m}+\frac{i}{8}\big(4\d^{[\m}_\n\d^{\r]}_\s+i\g^5 \e^\m{}_{\n}{}^{\r}{}_\s\big) \g^\n\g^5\cd_\r\Big(\ve\frac{\d}{\d C_\s}\Big)+\frac{3i}{4}\g^5\h(\ve)\frac{\d}{\d C_\m}+\g_\n\ve\frac{\d}{\d B_{\n\m}}\NO\\
&+\frac14\g^5\g^\m\big(\g^{\r\s} f_{\r\s}+\g^5D^I\big)\ve\frac{\d}{\d D^I},\NO\\
\rule{.0cm}{1.cm}\d_\ve\Big(\frac{\d}{\d\lbar\l^I}\Big)
=&\;\frac12\g_\m\ve\frac{\d}{\d a_\m^I}-\g^5\g^\m\cd_\m\Big(\ve\frac{\d}{\d D^I}\Big),
\eal
where $\h(\ve)$ is given in \eqref{CtoNM} and we have neglected terms of the schematic form $\j\frac{\d}{\d\lbar\j}$, $\j\frac{\d}{\d \lbar\l^I}$ and $\lbar\j\j\frac{\d}{\d D^I}$ in the transformation of the supercurrent. Notice that the supersymmetry transformations \eqref{operator-trans-cl} of the functional derivatives are directly related with the l.h.s. of the supercurrent conservation Ward identity in \eqref{NM-WardIDs-flavors}.

The full supersymmetry transformations of the fermionic operators in the quantum theory are
\bal\label{operator-trans}
\d_\ve\<\Hat\cs^\m\>_s
=&\;\frac{1}{2}\g^a\ve\<\Hat\ct^\m_a\>_s+\frac{i}{8}\big(4\d^{[\m}_\n\d^{\r]}_\s+i\g^5 \e^\m{}_{\n}{}^{\r}{}_\s\big) \g^\n\g^5\cd_\r\big(\ve\<\Hat\cj^\s\>_s\big)+\frac{3i}{4}\g^5\h(\ve)\<\Hat\cj^\m\>_s+\g_\n\ve\<\Hat\ck^{\n\m}\>\NO\\
&+\frac14\g^5\g^\m\big(\g^{\r\s} f_{\r\s}+\g^5D^I\big)\ve\<\co_I\>_s+\sum_{i=1}^8\k^{(i)}_{\{IJ\ldots\}} \S^{(i)\{IJ\ldots\}\m}(\ve)+\S^\m_{ct}(\ve),\NO\\
\rule{.0cm}{.7cm}\d_\ve\<\c_I\>_s
=&\;\frac12\g_\m\ve\<j^\m_I\>_s-\g^5\g^\m\cd_\m\big(\ve\<\co_I\>_s\big)+\sum_{i=3}^8\k^{(i)}_{\{IJ\ldots\}} \X^{(i)\{J\ldots\}}(\ve)+\X_I^{ct}(\ve),
\eal
where again we have neglected terms of the schematic form $\j\<\Hat\cs\>_s$, $\j\<\c_I\>_s$ and $\lbar\j\j\<\co_I\>_s$ in the transformation of the supercurrent. The anomalous contributions $\S^{(i)\{IJ\ldots\}\m}(\ve)$ and $\X^{(i)\{J\ldots\}}(\ve)$ to these transformations, as well as the contributions $\S^\m_{ct}(\ve)$ and $\X_I^{ct}(\ve)$ due to the counterterms \eqref{scheme-ct}, are obtained by evaluating the derivatives of the supersymmetry anomaly \eqref{NM-anomalies-vector-fermionic} with respect to the gravitino and the flavorinos using the expressions \eqref{NM-fermionic-anomalies}, \eqref{NM-anomalies-fermionic-vector-components} and \eqref{scheme-AQ}:
\bal
\rule{.0cm}{.7cm}\S^{(1)\m}(\ve)=&\;\frac{i}{2}\big(4\d^{[\m}_\n\d^{\r]}_\s+i\g^5 \e^\m{}_{\n}{}^{\r}{}_\s\big) \g^\n\g^5\cd_\r\big(\ve\;\wt G^{\s\k}C_\k\big)+3i\g^5\h(\ve)\wt G^{\m\n}C_\n+\frac{9}{2}\cd_\n\big(\wt G^{\m\n}\h(\ve)\big)\NO\\
&\;-\frac{9i}{4}\big(\g^{[\m}{}_\r\d^{\n]}_\s-\d^{[\m}_\r\d^{\n]}_\s\big)\g^5\cd_\n\big(G^{\r\s}\h(\ve)\big)-\frac{81}{8}\cd_\n\big(P_{\r\s}g^{\r[\s}\g^{\m\n]}\h(\ve)\big)+\co(\j^2),\NO\\
\rule{.0cm}{.7cm}\S^{(2)\m}(\ve)=&\;-4\de_{\r}\big(C_{\s}\wt R^{\s\l\r\k}\big)\d^\m_{(\k}\g_{\l)}\ve-G_{\r\s}\wt R^{\r\s\m\n}\g_\n\ve-10i\big(\g^{[\m}{}_\r\d^{\n]}_\s-\d^{[\m}_\r\d^{\n]}_\s\big)\g^5\cd_\n\big(G^{\r\s}\h(\ve)\big)\NO\\
&\;-9\cd_\n\big(P_{\r\s}g^{\r[\s}\g^{\m\n]}\h(\ve)\big)+3\cd_\n\Big[\Big(R^{\m\n\r\s}\g_{\r\s}-\frac12Rg_{\r\s}g^{\r[\s}\g^{\m\n]}\Big)\h(\ve)\Big]+\co(\j^2),\NO\\
\rule{.0cm}{.7cm}\S^{(3)I\m}(\ve)=&\;-\frac{4i}{3}\big(4\d^{[\m}_\n\d^{\r]}_\s+i\g^5 \e^\m{}_{\n}{}^{\r}{}_\s\big) \g^\n\g^5\cd_\r\big(\ve\;\wt G^{\s\k}a^I_\k\big)-8i\g^5\h(\ve)\wt G^{\m\n}a^I_\n-8\cd_\n\big(\wt f^{I\m\n}\h(\ve)\big)\NO\\
&\;-4\de_{\r}\big(a^I_{\s}\wt R^{\s\l\r\k}\big)\d^\m_{(\k}\g_{\l)}\ve-f^I_{\r\s}\wt R^{\r\s\m\n}\g_\n\ve+\co(\j^2,\j\l,\l^2),\NO\\
\rule{.0cm}{.7cm}\S^{(4)I\m}(\ve)=&\;i\Big(\cd_\n-\frac{i}{2}V^\t\g^5\g_\t\g_\n\Big)\big(\wt F^{\r\s}a_\r^I\g^5\g^{\m\n}\g_\s\ve\big)+\co(\j^2,\j\l,\l^2),\NO\\
\rule{.0cm}{.7cm}\S^{(5)(IJ)\m}(\ve)=&\;\co(\j^2,\j\l,\l^2),\NO\\
\rule{.0cm}{.7cm}\S^{(6)(IJK)\m}(\ve)=&\;\co(\j^2,\j\l,\l^2),\NO\\
\rule{.0cm}{.7cm}\S^{(7)I\m}(\ve)=&\;i\e^{\m\n\r\s}A_\r a_\s^I\g_\n\ve+\frac{i}{2}\Big(\cd_\n-\frac{i}{2}V^\t\g^5\g_\t\g_\n\Big)\big(a_\s^I\g^{\m\n}\g^\s\ve\big)+\co(\j^2,\j\l,\l^2),\NO\\
\rule{.0cm}{.7cm}\S^{(8)[IJ]\m}(\ve)=&\;\frac{i}{2}\e^{\m\n\r\s}a_\r^Ia_\s^J\g_\n\ve+\co(\j^2,\j\l,\l^2),
\eal
\bal
\S^\m_{ct}(\ve)=&\;\frac{i}{2}\a^{(4)}_{I}\Big(\cd_\n-\frac{i}{2}V^\t\g^5\g_\t\g_\n\Big)\big[\big(\wt f^{I\r\s}A_\r-2\wt F^{\r\s}a_\r^I\big)\g^5\g^{\m\n}\g_\s\ve\big]\NO\\
&\;+\frac{i}{2}\a^{(5)}_{(IJ)}\Big(\cd_\n-\frac{i}{2}V^\t\g^5\g_\t\g_\n\Big)\big(\wt f^{(I\r\s}a_\r^{J)}\g^5\g^{\m\n}\g_\s\ve\big)+\co(\j^2,\j\l,\l^2),
\eal
\bal
\rule{.0cm}{.7cm}\X^{(3)}(\ve)=&\;\co(\j^2,\j\l,\l^2),\NO\\
\rule{.0cm}{.7cm}\X^{(4)}(\ve)=&\;\co(\j^2,\j\l,\l^2),\NO\\
\rule{.0cm}{.7cm}\X^{(5)I}(\ve)=&\;-2A^\m\wt f^{I}_{\m\n}\g^\n\ve+\co(\j^2,\j\l,\l^2),\NO\\
\rule{.0cm}{.7cm}\X^{(6)(IJ)}(\ve)=&\;-2\wt f^{(I\m\n}a^{J)}_\m\g_\n\ve+\co(\j^2,\j\l,\l^2),\NO\\
\rule{.0cm}{.7cm}\X^{(7)}(\ve)=&\;A_\m\g^5\g^\m\ve+\co(\j^2,\j\l,\l^2),\NO\\
\rule{.0cm}{.7cm}\X^{(8)I}(\ve)=&\;-a^I_\m\g^5\g^\m\ve+\co(\j^2,\j\l,\l^2),
\eal
\bal
\X_I^{ct}(\ve)=&\;=-\a^{(4)}_{I}\wt F^{\m\n}A_\m\g_\n\ve+\a^{(5)}_{(IJ)}\big(2A^\m\wt f^{J}_{\m\n}\g^\n\ve-\wt F^{\m\n}a_\m^{J}\g_\n\ve\big)+\co(\j^2,\j\l,\l^2).
\eal

Notice that most of these terms are to leading order independent of the fermionic fields and therefore lead to an anomalous transformation for the fermionic operators on {\em purely bosonic} backgrounds. This has important implications for supersymmetric theories on purely bosonic backgrounds that admit new minimal Killing spinors, as we briefly discuss in the next section.

\section{Supersymmetric backgrounds and rigid supersymmetry anomalies}
\label{rigid}

A notion of rigid supersymmetry exits on purely bosonic backgrounds of new minimal supergravity for which the Killing spinor equations  
\bal\label{new-minimal-KSEs}
\d\j_\m=&\;\cd_\m\ve_o+\frac{i}{2}V^\n\g_\m\g_\n\g^5\ve_o=0,\NO\\
\d\l^I=&\;-\frac14\big(\g^{\m\n} f^I_{\m\n}+\g^5D^I\big)\ve_o=0,
\eal
admit non trivial solutions $\ve_o$. Note that in the Killing spinor equations $\ve_0$ is taken to be a $c$-number commuting spinor that transforms trivially under the symmetries of new minimal supergravity, in contrast to the local supersymmetry parameter $\ve$ that is Grassmann-valued and transforms according to \eqref{param-trans}. Moreover, the fact that the supersymmetry transformation of the gravitino in new minimal supergravity coincides with a combined $\cq$- and $\cs$-supersymmetry transformation in $\cn=1$ conformal supergravity with composite gauge field $C_\m=A_\m-\frac32 V_\m$ and $\cs$-supersymmetry parameter as in \eqref{CtoNM} implies that {\em locally}, supersymmetric backgrounds of new minimal and conformal supergravity coincide. However, non trivial solutions of the new minimal Killing spinor equations \eqref{new-minimal-KSEs} are nowhere vanishing, while those of conformal supergravity may have zeros \cite{Dumitrescu:2012ha}. Hence, globally, new minimal Killing spinors are also Killing spinors of conformal supergravity, but only a subset of conformal supergravity Killing spinors correspond to global Killing spinors of new minimal supergravity.  

Supersymmetric backgrounds of various off-shell supergravities and in different dimensions (including new minimal and conformal supergravity backgrounds in four dimensions) have been studied extensively  \cite{Samtleben:2012gy,Klare:2012gn,Dumitrescu:2012ha,Liu:2012bi,Dumitrescu:2012at,Kehagias:2012fh,Closset:2012ru,Samtleben:2012ua,Cassani:2012ri,deMedeiros:2012sb,Kuzenko:2012vd,Hristov:2013spa,Pan:2013uoa,Imamura:2014ima,Alday:2015lta} (see also \cite{Blau:2000xg} for earlier work). The notion of rigid supersymmetry such backgrounds admit enables the non perturbative calculation of certain quantum field theory observables using supersymmetric localization techniques \cite{Pestun:2007rz} (see \cite{Pestun:2016zxk} for a comprehensive review). These techniques rely on the existence of a bosonic ``localizing'' operator that is $Q$-exact, i.e. it can be expressed as the supersymmetry variation of a fermionic operator. However, in order for the localization argument to hold, the $Q$-exactness of the localizing operator must be preserved at the quantum level. Supersymmetry anomalies can potentially spoil this property, thus invalidating the localization argument. 

As a concrete example, let us consider the transformation of the fermionic operators in the $R$-multiplet and flavor multiplets under the rigid supersymmetry associated with a new minimal Killing spinor $\ve_o$. The local supersymmetry transformations \eqref{operator-trans} imply that the corresponding rigid supersymmetry transformations take the form 
\bal\label{operator-trans-rigid}
\d_{\ve_o}\<\Hat\cs^\m\>
=&\;\frac{1}{2}\g^a\ve_o\<\Hat\ct^\m_a\>+\frac{i}{8}\big(4\d^{[\m}_\n\d^{\r]}_\s+i\g^5 \e^\m{}_{\n}{}^{\r}{}_\s\big) \g^\n\g^5\cd_\r\big(\ve_o\<\Hat\cj^\s\>\big)+\frac{3i}{4}\g^5\h(\ve_o)\<\Hat\cj^\m\>+\g_\n\ve_o\<\Hat\ck^{\n\m}\>\NO\\
&+\sum_{i=1}^8\k^{(i)}_{\{IJ\ldots\}} \S^{(i)\{IJ\ldots\}\m}(\ve_o)+\S^\m_{ct}(\ve_o),\NO\\
\rule{.0cm}{.7cm}\d_{\ve_o}\<\c_I\>
=&\;\frac12\g_\m\ve_o\<j^\m_I\>-\g^5\g^\m\cd_\m\big(\ve_o\<\co_I\>\big)+\sum_{i=3}^8\k^{(i)}_{\{IJ\ldots\}} \X^{(i)\{J\ldots\}}(\ve_o)+\X_I^{ct}(\ve_o),
\eal
where we have removed the subscript $s$ from the one-point functions to indicate that these are now expectations values on a specific background. Notice that the term proportional to the expectation value of the scalar operators $\co_I$ in the rigid supersymmetry transformation of the supercurrent vanishes due to the Killing spinor equations. The terms $\S^{(i)\{IJ\ldots\}\m}(\ve_o)$, $\S^\m_{ct}(\ve_o)$, $\X^{(i)\{J\ldots\}}(\ve_o)$ and $\X_I^{ct}(\ve_o)$ that originate in the supersymmetry anomaly \eqref{NM-anomalies-vector-fermionic} are local functions of the {\em bosonic} background and they are non vanishing on generic backgrounds that admit new minimal Killing spinors. In fact, the term $\S^{(1)\m}(\ve_o)$ corresponding to the $\k^{(1)}$ cocycle has been evaluated explicitly on a class of backgrounds that admit two real supercharges of opposite $R$-charge and was shown to be non zero 
\cite{Papadimitriou:2017kzw}. The presence of these terms in the rigid supersymmetry transformation of the fermionic operators implies that the linear combination of bosonic operators on the r.h.s. of the transformations \eqref{operator-trans-rigid} are not $Q$-exact, as one would expect based on the classical supersymmetry algebra. 

The rigid supersymmetry algebra deformation due to the supersymmetry anomaly has implications for supersymmetric observables on such backgrounds. An immediate consequence is that the BPS relation that the conserved charges of supersymmetric states satisfy is modified \cite{Papadimitriou:2017kzw}. The dependence of supersymmetric partition functions on the background is also affected. The {\em classical} $Q$-exactness of the linear combination of bosonic currents on the r.h.s. of the supercurrent transformation in \eqref{operator-trans-rigid} implies that supersymmetric partition functions do not depend on certain deformations of the supersymmetric background  \cite{Closset:2013vra,Closset:2014uda,Assel:2014paa}. This result was contradicted by a holographic computation in \cite{Genolini:2016ecx} that explicitly examined the dependence of the holographic partition function on deformations of the supersymmetric background (see also \cite{BenettiGenolini:2017zmu,BenettiGenolini:2018iuy}). The resolution to this  contradiction was provided in \cite{Papadimitriou:2017kzw}, where it was shown that the dependence of the partition function on the supersymmetric background is entirely due to the deformation of the supersymmetry algebra by the term $\S^{(1)\m}(\ve_o)$ coming from the supersymmetry anomaly.  

An interesting question in this context is whether the anomalous terms in the rigid supersymmetry transformation of the supercurrent can be removed by a local counterterm. To answer this question one should keep in mind that in the presence of an $R$-symmetry and/or flavor anomaly the commutator \eqref{susysusy-extended} implies that the supersymmetry anomaly \eqref{NM-anomalies-vector-fermionic} cannot be removed by a local counterterm without breaking diffeomorphism and/or local Lorentz symmetry. It follows that any local counterterm that can potentially remove the anomaly from the rigid supersymmetry transformation of the supercurrent will necessarily break diffeomorphism and/or local Lorentz invariance. However, an interesting scenario is that the required local counterterm only breaks the subset of diffeomorphisms that would break the classical supersymmetry invariance of the background. This scenario is realized in an analogous situation for supersymmetric Chern-Simons theories on Seifert manifolds in connection with the framing anomaly \cite{Imbimbo:2014pla}.
For supersymmetric backgrounds of the form $S^1\times M_3$ with $M_3$ a Seifert manifold, the local counterterm that eliminates the term $\S^{(1)\m}(\ve_o)$ in the transformation of the supercurrent should coincide with the counterterm used in \cite{Genolini:2016ecx}. It would be interesting to generalize this counterterm to the other anomaly cocycles that contribute to the supersymmetry anomaly \eqref{NM-anomalies-vector-fermionic}. 

\section{Discussion}
\label{discussion}

In this paper we have extended our earlier results for $\cn=1$ conformal supergravity \cite{Katsianis:2019hhg,Papadimitriou:2019gel} to non conformal theories with an arbitrary number of Abelian flavor symmetries. As anticipated, both $R$-symmetry and flavor symmetry anomalies lead to a supersymmetry anomaly, even in non conformal theories. This anomaly is cohomologically non trivial and cannot be removed by a local counterterm without breaking diffeomorphism and/or local Lorentz symmetry.  

It would be very interesting to generalize these results to non Abelian $R$-symmetry anomalies in theories with extended supersymmetry, as well as non Abelian flavor symmetries. Moreover, in 2, 6 and 10 dimensions one could consider the effect of gravitational anomalies that are also known to generate a supersymmetry anomaly \cite{Howe:1985uy,Itoyama:1985ni,Tanii:1985wy}. 

Another question to address is if and how supersymmetry anomalies are manifest in superspace. As we briefly discussed in section \ref{nm-anomalies}, the auxiliary fields in the superspace formulation of background supergravity act as symmetry compensators \cite{Shamir:1992ff}, which implies that the non trivial solutions of the Wess-Zumino consistency conditions in superspace and in components may not coincide. It is therefore desirable to clarify if there is any connection between the supersymmetry anomalies we found here and the superspace cocycles found in \cite{Bonora:1984pn,Bonora:2013rta} and \cite{Brandt:1993vd,Brandt:1996au}.

In section \ref{supercurrent} we saw that the supersymmetry anomaly in the conservation of the supercurrent implies that both the supercurrent and the fermionic operators in the flavor multiplets acquire an anomalous supersymmetry transformation. When restricted to bosonic backgrounds that admit Killing spinors, this implies that these operators transform anomalously under rigid supersymmetry, which has implications for supersymmetric quantum field theory observables on such backgrounds. Specifically, the supersymmetry algebra gets deformed, the BPS relation that the bosonic conserved charges characterizing supersymmetric states satisfy is modified, and the $Q$-exactness of localizing operators used in supersymmetric localization computations may not hold at the quantum level. It is therefore important to further understand the consequences of the supersymmetry anomaly in this context. In particular, it would be very interesting to understand to what extend the rigid supersymmetry anomaly can be eliminated by a local non covariant counterterm. This question should be addressed separately for each of the eight non trivial cocycles that contribute to the supersymmetry anomaly and for each class of supersymmetric backgrounds preserving a given number of supercharges. We hope to address some of these questions in future work.

\section*{Acknowledgments}

I would like to thank Loriano Bonora, Friedemann Brandt, Cyril Closset, Parameswaran Nair, Peter West and especially Piljin Yi for constructive comments and email correspondence. I also thank George Katsianis, Kostas Skenderis and Marika Taylor for collaboration on related work.

\appendix

\renewcommand{\theequation}{\Alph{section}.\arabic{equation}}

\setcounter{section}{0}

\section*{Appendix}
\setcounter{section}{0}

\section{Review of supersymmetry anomalies in $\cn=1$ conformal supergravity}
\label{CS-review}
\setcounter{equation}{0}

In this appendix we summarize the local symmetry algebra and quantum anomalies of  $\cn=1$ off-shell conformal supergravity in four dimensions obtained in \cite{Papadimitriou:2019gel}. The field content of $\cn=1$ conformal supergravity  \cite{Kaku:1977pa,Kaku:1977rk,Kaku:1978nz,Townsend:1979ki} (see \cite{VanNieuwenhuizen:1981ae,deWit:1981vgr,deWit:1983qkc,Fradkin:1985am} and chapter 16 of \cite{Freedman:2012zz} for pedagogical reviews) consists of the vielbein $e^a_\m$, an Abelian gauge field $C_\m$, and a Majorana gravitino $\j_{\m}$, comprising 5+3 bosonic and 8 fermionic off-shell degrees of freedom. Throughout this paper we denote the gauge field of conformal supergravity by $C_\m$ and its fieldstrength by $G_{\m\n}=\pa_\m C_\n-\pa_\n C_\m$, reserving $A_\m$ and $F_{\m\n}=\pa_\m A_\n-\pa_\n A_\m$ for the gauge field of new minimal supergravity. 

$\cn=1$ conformal supergravity can be constructed as a gauge theory of the superconformal algebra. In this construction $\cq$- and $\cs$-supersymmetry are on the same footing with corresponding gauge fields $\j_\m$ and $\f_\m$. The curvature constraints of $\cn=1$ conformal supergravity, however, imply that $\f_\m$ is not an independent field and is locally expressed in terms of the gravitino as   
\be\label{phi}
\f_\m\equiv \frac13\g^\n\Big(\cd_\n\j_\m-\cd_\m\j_\n-\frac{i}{2}\g^5 \e_{\n\m}{}^{\r\s}\cd_\r\j_\s\Big)=-\frac16\big(4\d^{[\r}_\m\d^{\s]}_\n+i\g^5 \e_{\m\n}{}^{\r\s}\big)\g^\n \cd_\r\j_\s,
\ee
where the covariant derivative acts on $\j_\m$ and $\f_\m$ as 
\bal\label{covD-psi-phi}
\cd_\m\j_\n\equiv&\;\Big(\pa_\m+\frac14\o_\m{}^{ab}(e,\j)\g_{ab}+i\g^5C_\m\Big)\j_\n-\G^\r_{\m\n}\j_\r\equiv \big(\mathscr{D}_\m+i\g^5C_\m\big)\j_\n,\NO\\
\cd_\m\f_\n=&\;\Big(\pa_\m+\frac14\o_\m{}^{ab}(e,\j)\g_{ab}-i\g^5C_\m\Big)\f_\n-\G^\r_{\m\n}\f_\r=\big(\mathscr{D}_\m-i\g^5C_\m\big)\f_\n,
\eal
with the spin connection given by 
\be
\o_\m{}^{ab}(e,\j)\equiv\o_\m{}^{ab}(e)+\frac14\big(\lbar\j_a\g_\m\j_b+\lbar\j_\m\g_a\j_b-\lbar\j_\m\g_b\j_a\big).
\ee
 $\o_\m{}^{ab}(e)$ denotes the unique torsion-free spin connection.

\subsection{Local symmetry transformations}

Besides diffeomorphisms $\x^\m(x)$, local frame rotations $\l^{ab}(x)$, U(1)$_R$ gauge transformations $\th(x)$, and $\cq$-supersymmetry transformations $\ve(x)$, the local algebra of $\cn=1$ conformal supergravity contains also Weyl and $\cs$-supersymmetry transformations, parameterized respectively by $\s(x)$ and $\h(x)$. The corresponding transformations of the $\cn=1$ conformal supergravity fields are 
\bal\label{sugra-trans}
\d e^a_\m=&\;\x^\l\pa_\l e^a_\m+e^a_\l\pa_\m\x^\l-\l^a{}_b e^b_\m+\s e^a_\m-\frac12\lbar\j_\m\g^a\ve,\NO\\
\d\j_\m=&\;\x^\l\pa_\l\j_\m+\j_\l\pa_\m\x^\l-\frac14\l_{ab}\g^{ab}\j_\m+\frac12\s\j_\m+\cd_\m\ve-\g_\m\h- i\g^5\th\j_\m,\NO\\
\d C_\m=&\;\x^\l\pa_\l C_\m+C_\l\pa_\m\x^\l+\frac{3i}{4}\lbar\f_\m\g^5\ve-\frac{3i}{4}\lbar\j_\m\g^5\h+\pa_\m\th.
\eal
Moreover, the quantity $\f_\m$ transforms as
\be\label{phi-trans}
\d\f_\m=\x^\l\pa_\l\f_\m+\f_\l\pa_\m\x^\l-\frac14\l_{ab}\g^{ab}\f_\m-\frac12\s\f_\m+\frac12\Big(P_{\m\n}+\frac{2i}{3}G_{\m\n}\g^5-\frac{1}{3}\wt G_{\m\n}\Big)\g^\n\ve+\cd_\m\h+i\g^5\th\f_\m,
\ee
where
\be\label{Schouten}
P_{\m\n}\equiv\frac12\Big(R_{\m\n}-\frac16Rg_{\m\n}\Big),
\ee
denotes the Schouten tensor in four dimensions and the dual fieldstrength $\wt G_{\m\n}$ is defined as 
\be\label{dualF}
\wt G_{\m\n}\equiv\frac12 \e_{\m\n}{}^{\r\s}G_{\r\s}.
\ee
The covariant derivatives of the spinor parameters $\ve$ and $\h$ are given respectively by
\bal\label{covD-parameters}
\cd_\m\ve\equiv&\;\Big(\pa_\m+\frac14\o_\m{}^{ab}(e,\j)\g_{ab}+i\g^5C_\m\Big)\ve\equiv \big(\mathscr{D}_\m+i\g^5C_\m\big)\ve,\NO\\
\cd_\m\h\equiv&\;\Big(\pa_\m+\frac14\o_\m{}^{ab}(e,\j)\g_{ab}-i\g^5C_\m\Big)\h\equiv \big(\mathscr{D}_\m-i\g^5C_\m\big)\h.
\eal
 
\subsection{Local symmetry algebra}

The symmetry algebra is determined by the commutators $[\d_{\O_{\rm C}},\d_{\O'_{\rm C}}]$ between any two of the transformations \eqref{sugra-trans} with the local parameters $\O_{\rm C}=(\s,\x,\l,\th,\ve,\h)$ of $\cn=1$ conformal supergravity. 
In order for the algebra to close off-shell the local parameters should also transform under the local symmetries according to
\bal\label{param-trans}
&\d\x^\m=\x'^\n\pa_\n\x^\m-\x^\n\pa_\n\x'^\m,\qquad 
\d\l^a{}_b=\x^\m\pa_\m\l^a{}_b,\qquad
\d\s=\x^\m\pa_\m\s,\qquad
\d\th=\x^\m\pa_\m\th,\NO\\
&\d\ve=\x^\m\pa_\m\ve+\frac12\s\ve-\frac14\l_{ab}\g^{ab}\ve-i\th\g^5\ve,\qquad 
\d\h=\x^\m\pa_\m\h-\frac12\s\h-\frac14\l_{ab}\g^{ab}\h+i\th\g^5\h.
\eal
The only non vanishing commutators of the resulting local symmetry algebra are the following: 
\begin{alignat}{3}\label{CS-local-algebra}
&\;[\d_\x,\d_{\x'}]=\d_{\x''}, &&\;\x''^\m=\x^\n\pa_\n\x'^\m-\x'^\n\pa_\n\x^\m,\NO\\
&\;[\d_\l,\d_{\l'}]=\d_{\l''}, &&\;\l''^a{}_b=\l'^a{}_c\l^c{}_b-\l^a{}_c\l'^c{}_b,\rule{.0cm}{.5cm}\NO\\
&\;[\d_\ve,\d_\h]=\d_\s+\d_\l+\d_\th, &&\;\s=\frac12\lbar\ve\h,\quad  \l^a{}_b=-\frac12\lbar\ve\g^a{}_b\h,\quad \th=-\frac{3i}{4}\lbar\ve\g^5\h,\NO\\
&\;[\d_\ve,\d_{\ve'}]=\d_\x+\d_\l+\d_\th, \qquad &&\;\x^\m=\frac12\lbar\ve'\g^\m\ve,\quad \l^a{}_b=-\frac12(\lbar\ve'\g^\n\ve)\;\o_\n{}^a{}_b,\quad \th=-\frac12(\lbar\ve'\g^\n\ve)C_\n.
\end{alignat}
As for the new minimal supergravity algebra (see footnote \ref{effective-susy}), we have dropped a supersymmetry transformation on the r.h.s. of the commutator $[\d_\ve,\d_{\ve'}]$ that plays no role to leading order in the fermions.

\subsection{Ward identities and anomalies}

The current multiplet of a supersymmetric quantum field theory coupled to background $\cn=1$ conformal supergravity consists of the stress tensor $\ct^\m_a$, the $R$-symmetry current $\cj^\m$, and the supercurrent $\cs^\m$. These are the local operators sourced respectively by the vielbein $e^a_\m$, the gauge field $C_\m$, and the gravitino $\j_\m$. The local symmetry transformations of $\cn=1$ conformal supergravity \eqref{sugra-trans} lead to the superconformal Ward identities 
\bal
\label{WardIDs}
&e^a_\m\nabla_\n\<\ct^\n_a\>_s+\nabla_\n(\lbar\j_\m  \<\cs^\n\>_s)-\lbar\j_\n\overleftarrow \cd_\m \<\cs^\n\>_s-G_{\m\n}\<\cj^\n\>_s\NO\\
&\hspace{2.cm}+C_\m\big(\nabla_\n\<\cj^\n\>_s+i\lbar\j_\n \g^5\<\cs^\n\>_s\big)-\o_\m{}^{ab}\Big(e_{\n [a}\<\ct^\n_{b]}\>_s+\frac14\lbar\j_\n\g_{ab}\<\cs^\n\>_s\Big)=0,\NO\\
&e_{\m[a} \<\ct^\m_{b]}\>_s+\frac14\lbar\j_\m\g_{ab} \<\cs^\m\>_s=0,\NO\\
&e^a_\m\<\ct^\m_a\>_s+\frac12\lbar\j_\m \<\cs^\m\>_s=\ca_W,\NO\\
&\nabla_\m \<\cj^\m\>_s+i\lbar\j_\m\g^5 \<\cs^\m\>_s=\ca_R,\NO\\
&\cd_\m \<\cs^\m\>_s-\frac12\g^a\j_\m \<\ct^\m_a\>_s-\frac{3i}{4}\g^5\f_\m\<\cj^\m\>_s=\ca_Q,\NO\\
&\g_\m \<\cs^\m\>_s-\frac{3i}{4}\g^5\j_\m\<\cj^\m\>_s=\ca_{S},
\eal
where $\<\cdots\>_s$ denotes a correlation function in the presence of arbitrary sources and $\ca_W$, $\ca_R$, $\ca_Q$ and $\ca_S$ are quantum anomalies. 

In a scheme where the mixed axial-gravitational anomaly enters only in the conservation of the $R$-current (see e.g. eq.~(2.43) of \cite{Jensen:2012kj}), the Wess-Zumino consistency conditions determine the general form of the superconformal anomalies to be \cite{Papadimitriou:2019gel}
\bal\label{anomalies}
\ca_W=&\;\frac{c}{16\p^2}\Big(W^2-\frac{8}{3}G^2\Big)-\frac{a}{16\p^2} E+\co(\j^2),\NO\\
\ca_R=&\;\frac{(5a-3c)}{27\p^2}\;\wt G G+\frac{(c-a)}{24\p^2}\cp,\NO\\
\ca_Q=&\;-\frac{(5a-3c)i}{9\p^2}\wt G^{\m\n}C_\m\g^5\f_\n+\frac{(a-c)}{6\p^2}\nabla_\m\big(C_\r \wt R^{\r\s\m\n} \big)\g_{(\n}\j_{\s)}-\frac{(a-c)}{24\p^2}G_{\m\n} \wt R^{\m\n\r\s} \g_\r\j_\s+\co(\j^3),\NO\\
\ca_{S}=&\;\frac{(5a-3c)}{6\p^2}\wt G^{\m\n}\Big(\cd_\m-\frac{2i}{3}C_\m\g^5\Big)\j_{\n}+\frac{ic}{6\p^2} G^{\m\n}\big(\g_{\m}{}^{[\s}\d_{\n}^{\r]}-\d_{\m}^{[\s}\d_{\n}^{\r]}\big)\g^5\cd_\r\j_\s\NO\\
&+\frac{3(2a-c)}{4\p^2}P_{\m\n}g^{\m[\n}\g^{\r\s]}\cd_\r\j_\s+\frac{(a-c)}{8\p^2}\Big(R^{\m\n\r\s}\g_{\m\n}-\frac12Rg_{\m\n}g^{\m[\n}\g^{\r\s]}\Big)\cd_\r\j_\s+\co(\j^3),\hspace{-.06cm}
\eal
where $a$ and $c$ are the central charges of the superconformal algebra, normalized 
so that for free chiral and vector multiplets they are given respectively by \cite{Anselmi:1997am} 
\be
a=\frac{1}{48}(N_\c+9N_v),\qquad c=\frac{1}{24}(N_\c+3N_v).
\ee
Besides the Schouten tensor $P_{\m\n}$ defined in \eqref{Schouten} and the gauge field curvatures  
\be\label{gauge-curvatures}
G^2\equiv G_{\m\n}G^{\m\n},\qquad G\wt G\equiv \frac12 \e^{\m\n\r\s}G_{\m\n}G_{\r\s},
\ee
the superconformal anomalies are expressed in terms of the square of the Weyl tensor $W^2$, the Euler density $E$ and the Pontryagin density $\cp$. In terms of the Riemann tensor these take the form 
\bal\label{metric-curvatures}
W^2\equiv&\; W_{\m\n\r\s}W^{\m\n\r\s}=R_{\m\n\r\s}R^{\m\n\r\s}-2R_{\m\n}R^{\m\n}+\frac13R^2,\NO\\
E=&\;R_{\m\n\r\s}R^{\m\n\r\s}-4R_{\m\n}R^{\m\n}+R^2,\NO\\
\cp\equiv&\;\frac12\e^{\k\l\m\n}R_{\k\l\r\s}R_{\m\n}{}^{\r\s}=\wt R^{\m\n\r\s}R_{\m\n\r\s},
\eal
where the dual Riemann tensor is defined as
\be
\label{dualR}
\wt R_{\m\n\r\s}\equiv\frac12\e_{\m\n}{}^{\k\l}R_{\k\l\r\s}.
\ee
Notice that $\wt R_{\m\n\r\s}$ is not symmetric under exchange of the first and second pair of indices.

\section{Solving the Wess-Zumino conditions in the presence of flavor symmetries}
\label{WZ-proof}
\setcounter{equation}{0}

In this appendix we demonstrate that for each of the six flavor anomaly coefficients $\k^{(3)}_{I}$, $\k^{(4)}_{I}$, $\k^{(5)}_{(IJ)}$, $\k^{(6)}_{(IJK)}$, $\k^{(7)}_I$ and $\k^{(8)}_{[IJ]}$ the bosonic anomalies in \eqref{NM-anomalies-vector-bosonic} and the corresponding fermionic anomalies in \eqref{NM-anomalies-fermionic-vector-components} form a consistent solution (i.e. non trivial cocycle) of the Wess-Zumino conditions \eqref{WZ-condition} for new minimal supergravity coupled to Abelian flavor multiplets. The only non trivial consistency conditions that need to be checked in each case are $[\d_\ve,\d_\th]\mathscr{W}=0$, $[\d_\ve,\d_{\o}]\mathscr{W}=0$, and $[\d_{\ve},\d_{\ve'}]\mathscr{W}=(\d_\th+\d_\o)\mathscr{W}$ with $\th=-\frac12(\lbar\ve'\g^\m\ve) A_\m$, $\o^I=-\frac12(\lbar\ve'\g^\m\ve) a^I_\m$. We will explicitly compute these commutators, keeping only the leading non trivial order in the fermionic background $\j_\m$ and $\l^I$. Moreover, we assume that total derivative terms can be dropped.

\subsection{$\k^{(3)}_{I}$ cocycle}

\begin{flushleft}
\begin{tabular}{l}
\hline\hline
\raisebox{0.1cm}{\mbox{$[\d_\ve,\d_\th]\mathscr{W}=0$:\rule{0cm}{.5cm}\hspace{13.8cm}}}\\
\hline\hline
\end{tabular}
\end{flushleft}

The $R$-symmetry anomaly does not contain any term proportional to the anomaly coefficient $\k^{(3)}_{I}$, while the corresponding term in the supersymmetry anomaly is invariant under $R$-symmetry gauge transformations. Consequently, we trivially have 
\bal
\d_{\ve}\d_{\th} \mathscr{W}=&\;-\d_{\ve}\int d^4x\;e\left.\th\ca_{R}\right|_{\k^{(3)}_{I}}=0,\\\NO\\
\d_{\th} \d_{\ve}\mathscr{W}=&\;-\k^{(3)}_{I}\d_{\th}\int d^4x\;e\;\lbar\ve\ca_Q^{(3)I}=0,
\eal
which indeed give
\be
[\d_\ve,\d_\th]\mathscr{W}=0.
\ee

\begin{flushleft}
\begin{tabular}{l}
\hline\hline
\raisebox{0.1cm}{\mbox{$[\d_\ve,\d_{\o}]\mathscr{W}=0$:\rule{0cm}{.5cm}\hspace{13.7cm}}}\\
\hline\hline
\end{tabular}
\end{flushleft}

This commutator is similar to the commutator $[\d_\ve,\d_{\th}]\mathscr{W}=0$ for the gravity multiplet cocycles $\k^{(1)}$ and $\k^{(2)}$. We have,
\bal
\d_{\ve}\d_{\o} \mathscr{W}=&\;-\k^{(3)}_{I}\d_{\ve}\int d^4x\;e\;\o^I\Big(\cp-\frac83G\wt G\Big)\\
=&\;-\k^{(3)}_{I}\int d^4x\;e\;\o^I\e^{\m\n\r\s}\Big(\frac12\d_{\ve}(R_{\m\n\k\l}R^{\k\l}{}_{\r\s})-\frac{4}{3}\d_{\ve}(G_{\m\n}G_{\r\s})\Big)\NO\\
=&\;2\k^{(3)}_{I}\int d^4 x\;e\;\nabla_\r\big(\pa_\m\o^I\;\e^{\k\l\m\n}R^{\r\s}{}_{\k\l}\big)\lbar\ve\g_{(\s}\j_{\n)}\NO\\
&\;-4i\k^{(3)}_{I}\int d^4 x\;e\;\pa_\r\o^I\;\e^{\m\n\r\s}G_{\m\n}\lbar\ve\g^5\f_\s+4i\k^{(3)}_{I}\int d^4 x\;e\;\pa_\r\o^I\;\e^{\m\n\r\s}G_{\m\n}\lbar\h(\ve)\g^5\j_\s,\\\NO\\
\d_{\o} \d_{\ve}\mathscr{W}=&\;-\k^{(3)}_{I}\d_{\o}\int d^4x\;e\;\lbar\ve\ca_Q^{(3)I}\NO\\
=&\;2\k^{(3)}_{I}\int d^4 x\;e\;\nabla_\m\big(\pa_\r\o^I\e^{\r\s\k\l}  R_{\k\l}{}^{\m\n}\big)\lbar\ve\g_{(\n}\j_{\s)}\NO\\
&\;-4i\k^{(3)}_{I}\int d^4 x\;e\;\e^{\m\n\r\s}G_{\r\s}\pa_\m\o^I\lbar\ve\g^5\Big(\f_\n -\frac{i}{2}V^\k\g_\k\g^5\j_{\n}\Big),
\eal
where $\h(\ve)$ is given in \eqref{CtoNM}. Subtracting the two expressions gives
\be
[\d_\ve,\d_{\o}]\mathscr{W}=0.
\ee

\begin{flushleft}
\begin{tabular}{l}
\hline\hline
\raisebox{0.1cm}{\mbox{$[\d_{\ve},\d_{\ve'}]\mathscr{W}=\d_\o\mathscr{W}$ with $\o^I=-\frac12(\lbar\ve'\g^\m\ve) a^I_\m$:\rule{0cm}{.5cm}\hspace{9.cm}}}\\
\hline\hline
\end{tabular}
\end{flushleft}

This commutator is more involved, but it is closely related to the corresponding commutator for the gravity multiplet cocycles $\k^{(1)}$ and $\k^{(2)}$ upon replacing the flavor symmetry with $R$-symmetry. Two consecutive supersymmetry transformations give
\bal
\d_{\ve'} \d_{\ve}\mathscr{W}=&\;-\k^{(3)}_{I}\d_{\ve'}\int d^4x\;e\;\lbar\ve\ca_Q^{(3)I}\NO\\
=&\;-4i\k^{(3)}_{I}\int d^4 x\;e\;\e^{\m\n\r\s}G_{\m\n}a^I_\r\lbar\ve\g^5\d_{\ve'}\f_\s+4i\k^{(3)}_{I}\int d^4 x\;e\;\e^{\m\n\r\s}G_{\m\n}a^I_\r\lbar\h\g^5\d_{\ve'}\j_\s\NO\\
&\;-4\k^{(3)}_{I}\int d^4x\;e\;\e^{\m\n\r\s} f^{I}_{\r\s}\lbar\h\cd_\m\d_{\ve'}\j_{\n}\NO\\
&\;+2\k^{(3)}_{I}\int d^4x\;e\;\e^{\m\n\r\s}\nabla_\k\big(a^I_\r R^{\k\l}{}_{\m\n} \big)\lbar\ve\g_{(\l}\d_{\ve'}\j_{\s)}-\frac12\k^{(3)}_{I}\int d^4x\;e\;\e^{\m\n\r\s}f^I_{\r\s} R^{\k\l}{}_{\m\n} \lbar\ve\g_\k\d_{\ve'}\j_\l\NO\\
=&\;-4i\k^{(3)}_{I}\int d^4 x\;e\;\e^{\m\n\r\s}G_{\m\n}a^I_\r\lbar\ve\g^5\Big(\frac12P_{\s\l}+\frac{i}{3}G_{\s\l}\g^5-\frac{1}{12}\e_{\s\l}{}^{\k\t}G_{\k\t}\Big)\g^\l\ve'\NO\\
&\;-4i\k^{(3)}_{I}\int d^4 xe\;\e^{\m\n\r\s}G_{\m\n}a^I_\r\lbar\ve\g^5\cd_\s\h'\NO\\
&\;+4i\k^{(3)}_{I}\int d^4 x\;e\;\e^{\m\n\r\s}G_{\m\n}a^I_\r\lbar\h\g^5\cd_\s\ve'-4i\k^{(3)}_{I}\int d^4x\;e\;\e^{\m\n\r\s}G_{\m\n}a^I_\r\lbar\h\g^5\g_\s\h'\NO\\
&\;-4\k^{(3)}_{I}\int d^4x\;e\;\e^{\m\n\r\s}f^I_{\m\n}\lbar\h \cd_\r \cd_\s\ve'+4\k^{(3)}_{I}\int d^4x\;e\;\e_{\m\n}{}^{\r\s}f^{I\m\n}\lbar\h \g_\s \cd_\r\h'\NO\\
&\;-2\k^{(3)}_{I}\int d^4x\;e\;\e^{\m\n\r\s}a^I_\r R^{\k\l}{}_{\m\n} \nabla_\k(\lbar\ve\g_{(\l}\cd_{\s)}\ve')-2\k^{(3)}_{I}\int d^4x\;e\;\cancelto{0}{\e^{\m\n\r\s}\nabla_\k\big(a^I_\r R^{\k}{}_{\s\m\n} \big)}\lbar\ve\h'\NO\\
&\;-\frac12\k^{(3)}_{I}\int d^4x\;e\;\e^{\m\n\r\s}f^I_{\r\s} R^{\k\l}{}_{\m\n} \lbar\ve\g_\k \cd_\l\ve'+\frac12\k^{(3)}_{I}\int d^4x\;e\;\e^{\m\n\r\s}f^I_{\r\s} R^{\k\l}{}_{\m\n} \lbar\ve\g_{\k\l}\h',
\eal
where again $\h(\ve)$ is given in \eqref{CtoNM}.

We will first show that all terms involving $\h$ or $\h'$ sum to zero in the commutator $[\d_\ve,\d_{\ve'}]$. The term proportional to $\lbar\h\g^5\g_\s\h'$ does not contribute to the commutator since 
\be
\lbar\h\g^5\g_\s\h'=\lbar\h'\g^5\g_\s\h.
\ee
Moreover,
\bal
\lbar\h\g_\s \cd_\r\h'-\lbar\h'\g_\s \cd_\r\h=\de_\r(\lbar\h\g_\s\h').
\eal
Integrating by parts and using the Bianchi identity $\e^{\m\n\r\s}\pa_\r f^I_{\m\n}=0$ we therefore find that the term proportional to $\lbar\h\g_\s \cd_\r\h'$ does not contribute to the commutator either. Using the relations 
\be
-\lbar\ve\g^5\cd_\s\h'+\lbar\h\g^5\cd_\s\ve'-(-\lbar\ve'\g^5\cd_\s\h+\lbar\h'\g^5\cd_\s\ve)=\de_\s(\lbar\h\g^5\ve'-\lbar\ve\g^5\h'),
\ee
and 
\be
2\cd_{[\m} \cd_{\n]}\ve=\Big(\frac14 R_{\m\n\r\s}\g^{\r\s}+i\g^5G_{\m\n}\Big)\ve,
\ee
as well as the Bianchi identity $\e^{\m\n\r\s}\pa_\r G_{\m\n}=0$, the remaining terms involving $\h$ or $\h'$ combine into:
\bal
&2i\k^{(3)}_{I}\int d^4 xe\;\e^{\m\n\r\s}G_{\m\n}f^I_{\r\s}(\lbar\h\g^5\ve'-\lbar\ve\g^5\h')+\frac12\k^{(3)}_{I}\int d^4x\;e\;\e^{\m\n\r\s}f^I_{\r\s} R^{\k\l}{}_{\m\n} (\lbar\ve\g_{\k\l}\h'-\lbar\ve'\g_{\k\l}\h)\NO\\
&-2\k^{(3)}_{I}\int d^4x\;e\;\e^{\m\n\r\s}f^I_{\m\n}\Big(\frac14 R_{\r\s}{}^{\k\l}(\lbar\ve\g_{\k\l}\h'-\lbar\ve'\g_{\k\l}\h)+iG_{\r\s}(\lbar\h\g^5\ve'-\lbar\ve\g^5\h')\Big)=0.
\eal
Therefore, all terms involving $\h$ or $\h'$ sum to zero in the commutator $[\d_\ve,\d_{\ve'}]$.

From the remaining terms we get 
\bal
[\d_\ve,\d_{\ve'}]\mathscr{W}
=&\;\frac{8}{3}\k^{(3)}_{I}\int d^4 x\;e\;\e^{\m\n\r\s}G_{\m\n}a^I_\r G_{\s\l}\lbar\ve'\g^\l\ve\NO\\
&-\frac12\k^{(3)}_{I}\int d^4x\;e\;\e^{\m\n\r\s}a^I_\r R^{\k\l}{}_{\m\n} R_{\k\l\s\t}(\lbar\ve'\g^{\t}\ve)\NO\\
&-\k^{(3)}_{I}\int d^4x\;e\;\e^{\m\n\r\s}a^I_\r R^{\k\l}{}_{\m\n} \nabla_\k\nabla_{\s}(\lbar\ve'\g_{\l}\ve)\NO\\
&-\frac12\k^{(3)}_{I}\int d^4x\;e\;\e^{\m\n\r\s}f^I_{\r\s} R^{\k\l}{}_{\m\n} \de_\l(\lbar\ve'\g_\k \ve).
\eal
The last two terms can be rearranged as
\bal
&\hspace{-2.0cm}\e^{\m\n\r\s}a^I_\r R^{\k\l}{}_{\m\n}\nabla_\k\nabla_\s(\lbar\ve'\g_\l\ve)-\frac12\e^{\m\n\r\s}f^I_{\r\s} R^{\k\l}{}_{\m\n}\nabla_\k(\lbar\ve'\g_\l\ve)\NO\\
=&\;\e^{\m\n\r\s}a^I_\r R^{\k}{}_{\l\m\n}[\nabla_\k,\nabla_\s](\lbar\ve'\g^\l\ve)+\de_\s\big(\e^{\m\n\r\s}a^I_\r R^{\k\l}{}_{\m\n}\nabla_\k(\lbar\ve'\g_\l\ve)\big)\NO\\
=&\;\de_\s\big(\e^{\m\n\r\s}a^I_\r R^{\k\l}{}_{\m\n}\nabla_\k(\lbar\ve'\g_\l\ve)\big)+\e^{\m\n\r\s}a^I_\r R^{\k\l}{}_{\m\n}R_{\k\s\l\t}(\lbar\ve'\g^\t\ve)\NO\\
=&\;\de_\s\big(\e^{\m\n\r\s}a^I_\r R^{\k\l}{}_{\m\n}\nabla_\k(\lbar\ve'\g_\l\ve)\big)+\frac12\e^{\m\n\r\s}a^I_\r R^{\k\l}{}_{\m\n}R_{\k\l\s\t}(\lbar\ve'\g^\t\ve),
\eal
so that
\bal
[\d_\ve,\d_{\ve'}]\mathscr{W}=&\;\frac{8}{3}\k^{(3)}_{I}\int d^4 x\;e\;\e^{\m\n\r\s}G_{\m\n}a^I_\r G_{\s\l}\lbar\ve'\g^\l\ve\NO\\
&-\k^{(3)}_{I}\int d^4x\;e\;\e^{\m\n\r\s}a^I_\r R^{\k\l}{}_{\m\n} R_{\k\l\s\t}(\lbar\ve'\g^{\t}\ve).
\eal
In order to simplify these expressions we notice that for any antisymmetric tensor $\Th_{\m\n}$ and vector $\J_\m$ in four dimensions we have $\Th_{[\l\s}\Th_{\m\n}\J_{\r]}=0$, which leads to the identity
\be\label{5antisym}
\e^{\m\n\r\s}\Th_{\m\n}\Th_{\s\l}\J_\r=-\frac14\e^{\m\n\r\s}\Th_{\m\n}\Th_{\r\s}\J_\l.
\ee
In particular,
\bal\label{5-antisym-ids}
\e^{\m\n\r\s}G_{\m\n}G_{\s\l}a^I_\r=&\;-\frac14\e^{\m\n\r\s}G_{\m\n}G_{\r\s}a^I_\l,\NO\\
\e^{\m\n\r\s}R_{\m\n\k\l}R_{\s\t}{}^{\k\l}a^I_\r=&\;-\frac14\e^{\m\n\r\s}R_{\m\n\k\l}R_{\r\s}{}^{\k\l}a^I_\t,
\eal
and so we finally get 
\bal
[\d_\ve,\d_{\ve'}]\mathscr{W}=&\;-\frac{2}{3}\k^{(3)}_{I}\int d^4 x\;e\;\e^{\m\n\r\s}G_{\m\n}G_{\r\s}(a^I_\l\lbar\ve'\g^\l\ve)+\frac{1}{4}\k^{(3)}_{I}\int d^4x\;e\;\e^{\m\n\r\s}R^{\k\l}{}_{\m\n}R_{\k\l\r\s} (a^I_\t\lbar\ve'\g^\t\ve)\NO\\
=&\;-\k^{(3)}_{I}\int d^4 x\;e\;\o^I\ca^I,
\eal
with
\be
\o^I=-\frac12(\lbar\ve'\g^\l\ve)a^I_\l,
\ee
as required by the Wess-Zumino consistency conditions.

\subsection{$\k^{(4)}_I$ cocycle}

\begin{flushleft}
\begin{tabular}{l}
\hline\hline
\raisebox{0.1cm}{\mbox{$[\d_\ve,\d_\th]\mathscr{W}=0$:\rule{0cm}{.5cm}\hspace{13.8cm}}}\\
\hline\hline
\end{tabular}
\end{flushleft}

As for the $\k^{(3)}_I$ cocycle, this commutator is trivially satisfied since
\bal
\d_{\ve}\d_{\th} \mathscr{W}=&\;-\d_{\ve}\int d^4x\;e\left.\th\ca_{R}\right|_{\k^{(4)}_I}=0,\\\NO\\
\d_{\th} \d_{\ve}\mathscr{W}=&\;-\k^{(4)}_I\d_{\th}\int d^4x\;e\;\lbar\ve\ca_Q^{(4)I}=0.
\eal

\begin{flushleft}
\begin{tabular}{l}
\hline\hline
\raisebox{0.1cm}{\mbox{$[\d_\ve,\d_{\o}]\mathscr{W}=0$:\rule{0cm}{.5cm}\hspace{13.7cm}}}\\
\hline\hline
\end{tabular}
\end{flushleft}

This commutator is an example of the connection between the Chern-Simons forms and the supersymmetry anomaly discussed in subsection \ref{flavor-anomalies}. We have, 
\bal
\d_{\ve}\d_{\o} \mathscr{W}=&\;-\frac12\k^{(4)}_I\int d^4x\;e\;\o^I\e^{\m\n\r\s}\d_{\ve}(F_{\m\n} F_{\r\s})\NO\\
&\;=-2\k^{(4)}_I\int d^4x\;e\;\o^I\e^{\m\n\r\s}F_{\m\n} \pa_\r \d_{\ve}A_{\s}\NO\\
&\;=\frac{i}{2}\k^{(4)}_I\int d^4x\;e\;\pa_\r\o^I\e^{\m\n\r\s}F_{\m\n}\lbar\ve\g_\s\g^{\k\l}\g^5\Big(\cd_\k\j_\l+\frac{i}{2}V^\t\g_\k\g_\t\g^5\j_\l\Big),\\\NO\\
\d_{\o} \d_{\ve}\mathscr{W}=&\;-\k^{(4)}_I\d_{\o}\int d^4x\;e\;\lbar\ve\ca_Q^{(4)I}\NO\\
=&\;\frac{i}{2}\k^{(4)}_I\int d^4x\;e\;\pa_\r\o^I\e^{\m\n\r\s}F_{\m\n}\lbar\ve\g_\s\g^{\k\l}\g^5\Big(\cd_\k\j_\l+\frac{i}{2}V^\t\g_\k\g_\t\g^5\j_\l\Big).
\eal
Hence,
\be
[\d_\ve,\d_{\o}]\mathscr{W}=0.
\ee

\begin{flushleft}
\begin{tabular}{l}
\hline\hline
\raisebox{0.1cm}{\mbox{$[\d_{\ve},\d_{\ve'}]\mathscr{W}=\d_\o\mathscr{W}$ with $\o^I=-\frac12(\lbar\ve'\g^\m\ve) a^I_\m$:\rule{0cm}{.5cm}\hspace{9.cm}}}\\
\hline\hline
\end{tabular}
\end{flushleft}

This commutator can be evaluated most efficiently by noticing that two successive supersymmetry transformations of the generating function can be expressed in terms of two successive supersymmetry transformations of the $R$-symmetry gauge field. Namely, 
\bal
\d_{\ve'} \d_{\ve}\mathscr{W}=&\;-\k^{(4)}_I\d_{\ve'}\int d^4x\;e\;\lbar\ve\ca_Q^{(4)I}\NO\\
=&\;\k^{(4)}_I\int d^4x\;e\;\e^{\m\n\r\s}2F_{\m\n} a_\r^I\d_{\ve'}\d_{\ve}A_{\s}.
\eal

Using the local symmetry algebra \eqref{susysusy-extended} we therefore obtain
\bal
[\d_{\ve}, \d_{\ve'}]\mathscr{W}
=&\;\k^{(4)}_I\int d^4x\;e\;\e^{\m\n\r\s}2F_{\m\n} a_\r^I[\d_{\ve}, \d_{\ve'}]A_{\s}\NO\\
=&\;\k^{(4)}_I\int d^4x\;e\;\e^{\m\n\r\s}2F_{\m\n} a_\r^I(\d_\x+\d_\th)A_{\s}\NO\\
=&\;\;\k^{(4)}_I\int d^4x\;e\;\e^{\m\n\r\s}2F_{\m\n} a_\r^I(\x^\l\pa_\l A_\s+A_\k\pa_\s\x^\k+\pa_\s\th)\NO\\
=&\;\;\k^{(4)}_I\int d^4x\;e\;\e^{\m\n\r\s}2F_{\m\n} a_\r^I\x^\l F_{\l\s}\NO\\
=&\;-\int d^4x\;e\;\o^I\left.\ca_R\right|_{\k^{(4)}_{(IJ)}},
\eal
where 
\be
\x^\m=\frac12(\lbar\ve'\g^\m\ve),\qquad \th=-\frac12(\lbar\ve'\g^\m\ve) A_\m,\qquad \o^I=-\frac12(\lbar\ve'\g^\m\ve) a^I_\m,
\ee
and in the last step we have used the identity \eqref{5antisym}.

\subsection{$\k^{(5)}_{(IJ)}$ cocycle}

\begin{flushleft}
\begin{tabular}{l}
\hline\hline
\raisebox{0.1cm}{\mbox{$[\d_\ve,\d_\th]\mathscr{W}=0$:\rule{0cm}{.5cm}\hspace{13.8cm}}}\\
\hline\hline
\end{tabular}
\end{flushleft}

The Wess-Zumino commutation relations for the $\k^{(5)}_{(IJ)}$ cocycle are analogous to those of the $\k^{(4)}_{(IJ)}$ cocycle upon replacing the $R$-symmetry with the flavor symmetry gauge fields. We have, 
\bal
\d_{\ve}\d_{\th} \mathscr{W}=&\;-\frac12\k^{(5)}_{(IJ)}\int d^4x\;e\;\th\;\e^{\m\n\r\s}\d_{\ve}(f^I_{\m\n} f^J_{\r\s})\NO\\
&\;=-2\k^{(5)}_{(IJ)}\int d^4x\;e\;\th\;\e^{\m\n\r\s}f^I_{\m\n} \pa_\r \d_{\ve}a^J_{\s}\NO\\
&\;=\k^{(5)}_{(IJ)}\int d^4x\;e\;\pa_\r\th\;\e^{\m\n\r\s}f^I_{\m\n}\lbar\ve\g_\s\l^J,\\\NO\\
\d_{\th} \d_{\ve}\mathscr{W}=&\;-\k^{(5)}_{(IJ)}\d_{\th}\int d^4x\;e\;\lbar\ve\ca_Q^{(5)(IJ)}\NO\\
=&\;\k^{(5)}_{(IJ)}\int d^4x\;e\;\pa_\r\th\;\e^{\m\n\r\s}f^I_{\m\n}\lbar\ve\g_\s\l^J.
\eal
Hence,
\be
[\d_\ve,\d_\th]\mathscr{W}=0.
\ee

\begin{flushleft}
\begin{tabular}{l}
\hline\hline
\raisebox{0.1cm}{\mbox{$[\d_\ve,\d_{\o}]\mathscr{W}=0$:\rule{0cm}{.5cm}\hspace{13.7cm}}}\\
\hline\hline
\end{tabular}
\end{flushleft}

This commutator is again trivially satisfied since
\bal
\d_{\ve}\d_{\o} \mathscr{W}=&\;-\d_{\ve}\int d^4x\;e\left.\o^I\ca_{I}\right|_{\k^{(5)}_{(IJ)}}=0,\\\NO\\
\d_{\o} \d_{\ve}\mathscr{W}=&\;-\k^{(5)}_{(IJ)}\d_{\o}\int d^4x\;e\;\lbar\ve\ca_Q^{(5)(IJ)}=0.
\eal

\begin{flushleft}
\begin{tabular}{l}
\hline\hline
\raisebox{0.1cm}{\mbox{$[\d_{\ve},\d_{\ve'}]\mathscr{W}=\d_\th\mathscr{W}$ with $\th=-\frac12(\lbar\ve'\g^\m\ve) A_\m$:\rule{0cm}{.5cm}\hspace{9.2cm}}}\\
\hline\hline
\end{tabular}
\end{flushleft}

As for the $\k^{(4)}_{(IJ)}$ cocycle we evaluate this commutator by expressing two successive supersymmetry transformations of the generating function in terms of two successive supersymmetry transformations of the flavor gauge fields. Namely, 
\bal
\d_{\ve'} \d_{\ve}\mathscr{W}=&\;-\k^{(5)}_{(IJ)}\d_{\ve'}\int d^4x\;e\;\lbar\ve\ca_Q^{(5)(IJ)}\NO\\
=&\;\k^{(5)}_{(IJ)}\int d^4x\;e\;\e^{\m\n\r\s}2A_\m f^I_{\r\s}\d_{\ve'}\d_\ve a_\n^J.
\eal
Using the local symmetry algebra \eqref{susysusy-extended} this gives
\bal
[\d_{\ve}, \d_{\ve'}]\mathscr{W}=&\;\k^{(5)}_{(IJ)}\int d^4x\;e\;\e^{\m\n\r\s}2A_\m f^I_{\r\s}[\d_{\ve}, \d_{\ve'}] a_\n^J\NO\\
=&\;\k^{(5)}_{(IJ)}\int d^4x\;e\;\e^{\m\n\r\s}2A_\m f^I_{\r\s}(\d_\x+\d_\o) a_\n^J\NO\\
=&\;\k^{(5)}_{(IJ)}\int d^4x\;e\;\e^{\m\n\r\s}2A_\m f^I_{\r\s}(\x^\k\pa_\k a_\n^J+a_\k^J\pa_\n\x^\k+\pa_\n\o^I)\NO\\
=&\;\k^{(5)}_{(IJ)}\int d^4x\;e\;\e^{\m\n\r\s}2A_\m f^I_{\r\s}\x^\k f_{\k\n}^J\NO\\
=&\;\frac12\k^{(5)}_{(IJ)}\int d^4x\;e\;\e^{\m\n\r\s}f^I_{\m\n}f^J_{\r\s}\x^\k A_\k\NO\\
=&\;-\int d^4x\;e\;\th\left.\ca_R\right|_{\k^{(5)}_{(IJ)}},
\eal
where 
\be
\x^\m=\frac12(\lbar\ve'\g^\m\ve),\qquad \th=-\frac12(\lbar\ve'\g^\m\ve) A_\m,\qquad \o^I=-\frac12(\lbar\ve'\g^\m\ve) a^I_\m,
\ee
and in the last step we have used again the identity \eqref{5antisym}.

\subsection{$\k^{(6)}_{(IJK)}$ cocycle}

\begin{flushleft}
\begin{tabular}{l}
\hline\hline
\raisebox{0.1cm}{\mbox{$[\d_\ve,\d_\th]\mathscr{W}=0$:\rule{0cm}{.5cm}\hspace{13.8cm}}}\\
\hline\hline
\end{tabular}
\end{flushleft}

This commutator is trivially satisfied since
\bal
\d_{\ve}\d_{\th} \mathscr{W}=&\;-\d_{\ve}\int d^4x\;e\left.\th\ca_{R}\right|_{\k^{(6)}_{(IJK)}}=0,\\\NO\\
\d_{\th} \d_{\ve}\mathscr{W}=&\;-\k^{(6)}_{(IJK)}\d_{\th}\int d^4x\;e\;\lbar\ve\ca_Q^{(6)(IJK)}=0.
\eal

\begin{flushleft}
\begin{tabular}{l}
\hline\hline
\raisebox{0.1cm}{\mbox{$[\d_\ve,\d_{\o}]\mathscr{W}=0$:\rule{0cm}{.5cm}\hspace{13.7cm}}}\\
\hline\hline
\end{tabular}
\end{flushleft}

This commutator is also straightforward to evaluate:
\bal
\d_{\ve}\d_{\o} \mathscr{W}=&\;-\frac12\k^{(6)}_{(IJK)}\int d^4x\;e\;\o^I\e^{\m\n\r\s}\d_{\ve}(f^J_{\m\n}f^K_{\r\s})\NO\\
=&\;2\k^{(6)}_{(IJK)}\int d^4x\;e\;\pa_\r\o^I\e^{\m\n\r\s}f^J_{\m\n}\d_{\ve}a^K_{\s}\NO\\
=&\;\k^{(6)}_{(IJK)}\int d^4x\;e\;\pa_\r\o^I\e^{\m\n\r\s}f^J_{\m\n}\lbar\ve \g_\s\l^K,\\\NO\\
\d_{\o} \d_{\ve}\mathscr{W}=&\;-\k^{(6)}_{(IJK)}\d_{\o}\int d^4x\;e\;\lbar\ve\ca_Q^{(6)(IJK)}\NO\\
=&\;\k^{(6)}_{(IJK)}\int d^4x\;e\;\pa_\r\o^I\e^{\m\n\r\s}f^J_{\m\n}\lbar\ve \g_\s\l^K.
\eal
Hence,
\be
[\d_\ve,\d_{\o}]\mathscr{W}=0.
\ee

\begin{flushleft}
\begin{tabular}{l}
\hline\hline
\raisebox{0.1cm}{\mbox{$[\d_{\ve},\d_{\ve'}]\mathscr{W}=\d_\o\mathscr{W}$ with $\o^I=-\frac12(\lbar\ve'\g^\m\ve) a^I_\m$:\rule{0cm}{.5cm}\hspace{9.cm}}}\\
\hline\hline
\end{tabular}
\end{flushleft}

As for the $\k^{(4)}_{(IJ)}$ and $\k^{(5)}_{(IJ)}$ cocycles, this commutator can be evaluated by expressing two successive supersymmetry transformations of the generating function in terms of two successive supersymmetry transformations of the flavor gauge fields:
\bal
\d_{\ve'} \d_{\ve}\mathscr{W}=&\;-\k^{(6)}_{(IJK)}\d_{\ve'}\int d^4x\;e\;\lbar\ve\ca_Q^{(6)(IJK)}\NO\\
=&\;2\k^{(6)}_{(IJK)}\int d^4x\;e\;\e^{\m\n\r\s}a^I_\r f^J_{\m\n}\d_{\ve'}\d_\ve a_\s^K.
\eal
Hence,
\bal
[\d_{\ve},\d_{\ve'}]\mathscr{W}=&\;2\k^{(6)}_{(IJK)}\int d^4x\;e\;\e^{\m\n\r\s}a^I_\r f^J_{\m\n}[\d_{\ve},\d_{\ve'}] a_\s^K\NO\\
=&\;2\k^{(6)}_{(IJK)}\int d^4x\;e\;\e^{\m\n\r\s}a^I_\r f^J_{\m\n}(\d_\x+\d_\o) a_\s^K\NO\\
=&\;2\k^{(6)}_{(IJK)}\int d^4x\;e\;\e^{\m\n\r\s}a^I_\r f^J_{\m\n}\x^\k f^K_{\k\s}\NO\\
=&\;\frac12 \k^{(6)}_{(IJK)}\int d^4x\;e\;\x^\k a^I_\k\; \e^{\m\n\r\s} f^J_{\m\n}f^K_{\r\s}\NO\\
=&\;-\int d^4x\;e\;\o^I\left.\ca_I\right|_{\k^{(6)}_{(IJK)}},
\eal
where again 
\be
\x^\m=\frac12(\lbar\ve'\g^\m\ve),\qquad \o^I=-\frac12(\lbar\ve'\g^\m\ve) a^I_\m,
\ee
and we have once more made use of the identity \eqref{5antisym}.

\subsection{$\k^{(7)}_I$ cocycle}

\begin{flushleft}
\begin{tabular}{l}
\hline\hline
\raisebox{0.1cm}{\mbox{$[\d_\ve,\d_\th]\mathscr{W}=0$:\rule{0cm}{.5cm}\hspace{13.8cm}}}\\
\hline\hline
\end{tabular}
\end{flushleft}

Solving the Wess-Zumino conditions for the Fayet-Iliopoulos type cocycles $\k^{(7)}_I$ and $\k^{(8)}_{[IJ]}$ is somewhat more involved because their contribution to the $R$-symmetry and flavor anomalies contains terms quadratic in the fermions. We only outline the essential steps of this calculation here.

Checking the commutation relation $[\d_\ve,\d_\th]\mathscr{W}=0$ involves computing the supersymmetry transformation of the $R$-symmetry anomaly term proportional to $\k^{(7)}_I$. This requires a bit of algebra, but a sketch of the calculation is as follows:
\bal
\d_{\ve}\d_{\th} \mathscr{W}=&\;-\d_{\ve}\int d^4x\;e\left.\th\ca_{R}\right|_{\k^{(7)}_I}\NO\\
=&\;-\k^{(7)}_I\d_{\ve}\int d^4x\;e\;\th \big(D^I-i\e^{\m\n\r\s}a^I_\m \pa_\n B_{\r\s}+\lbar\l^I\g^5\g^\m\j_\m\big)\NO\\
=&\;-\k^{(7)}_I\int d^4x\;e\;\th\Big[\lbar\ve\g^5\g^\m\Big(\cd_\m\l^I+\frac14(\g^{\r\s} f^I_{\r\s}+\g^5D^I)\j_\m\Big)\NO\\
&\;-\frac{i}{2}\e^{\m\n\r\s}\pa_\n B_{\r\s}\lbar\ve\g_\m\l^I-i\e^{\m\n\r\s}\pa_\n(\lbar\ve\g_\r\j_\s)a^I_\m+\frac12\lbar\ve\g^\m\j_\m D^I\NO\\
&\;+\lbar\l^I\g^5\g^\m\Big(\cd_\m\ve+\frac{i}{2}V^\k\g_\m\g_\k\g^5\ve\Big)-\frac{1}{4}\lbar\j_\m\g^5\g^\m\big(\g^{\r\s}f_{\r\s}+\g^5 D^I\big)\ve\Big]\NO\\
=&\;\k^{(7)}_I\int d^4x\;e\;\pa_\m\th\; \lbar\ve\big(\g^5\g^\m\l^{I}+i\e^{\m\n\r\s}a_\n^{I}\g_\r\j_\s\big),\\\NO\\
\d_{\th} \d_{\ve}\mathscr{W}=&\;-\k^{(7)}_I\d_{\th}\int d^4x\;e\;\lbar\ve\ca_Q^{(7)I}\NO\\
=&\;\k^{(7)}_I\int d^4x\;e\;\pa_\m\th\; \lbar\ve\big(\g^5\g^\m\l^{I}+i\e^{\m\n\r\s}a_\n^{I}\g_\r\j_\s\big).
\eal
Hence, we arrive at the correct Wess-Zumino condition
\be
[\d_\ve,\d_\th]\mathscr{W}=0.
\ee

\begin{flushleft}
\begin{tabular}{l}
\hline\hline
\raisebox{0.1cm}{\mbox{$[\d_\ve,\d_{\o}]\mathscr{W}=0$:\rule{0cm}{.5cm}\hspace{13.7cm}}}\\
\hline\hline
\end{tabular}
\end{flushleft}

In order to verify this commutation relation we need to make use of the identity 
\bal
&\d_\ve\Big[e\Big(\e^{\m\n\r\s}A_\m \pa_\n B_{\r\s}-\frac{1}{2}R-3V_\m V^\m-\frac12\de_\n(\lbar\j^\n\g^\m\j_\m)+\frac{1}{2}\lbar\j_\m\g^{\m\r\s}\Big(\cd_\r\j_\s+\frac{3i}{4}V^\t\g_\r\g_\t\g^5\j_\s\Big)\Big)\Big]=\NO\\
&\pa_\n\Big[e\;\e^{\m\n\r\s}A_\m\lbar\ve\g_\r\j_\s+\frac12e\;\lbar\ve\g^\n\g^{\r\s}\Big(\cd_\r\j_\s+\frac{i}{2}V^\t\g_\r\g_\t\g^5\j_\s\Big)\Big].
\eal
The derivation of this identity is rather lengthy and we will not present it here. Given this identity, however, the commutation relation $[\d_\ve,\d_{\o}]\mathscr{W}=0$ follows trivially:
\bal
\d_{\ve}\d_{\o} \mathscr{W}=&\;-\d_{\ve}\int d^4x\;e\left.\o^I\ca_{I}\right|_{\k^{(7)}_I}\NO\\
=&\;-\k^{(7)}_I\d_\ve\int d^4x\;e\;\o^I\Big[i\e^{\m\n\r\s}A_\m \pa_\n B_{\r\s}-\frac{i}{2}R-3iV_\m V^\m-\frac{i}{2}\de_\n(\lbar\j^\n\g^\m\j_\m)\NO\\
&\;+\frac{i}{2}\lbar\j_\m\g^{\m\r\s}\Big(\cd_\r\j_\s+\frac{3i}{4}V^\t\g_\r\g_\t\g^5\j_\s\Big)\Big]\NO\\
=&\;\k^{(7)}_I\int d^4x\;e\;\pa_\n\o^I\Big[i\e^{\m\n\r\s}A_\m\lbar\ve\g_\r\j_\s+\frac{i}{2}\lbar\ve\g^\n\g^{\r\s}\Big(\cd_\r\j_\s+\frac{i}{2}V^\t\g_\r\g_\t\g^5\j_\s\Big)\Big],\\
\d_{\o} \d_{\ve}\mathscr{W}=&\;-\k^{(7)}_I\d_{\o}\int d^4x\;e\;\lbar\ve\ca_Q^{(7)I}\NO\\
=&\;\k^{(7)}_I\int d^4x\;e\;\pa_\n\o^I\Big[i\e^{\m\n\r\s}A_\m\lbar\ve\g_\r\j_\s+\frac{i}{2}\lbar\ve\g^\n\g^{\r\s}\Big(\cd_\r\j_\s+\frac{i}{2}V^\t\g_\r\g_\t\g^5\j_\s\Big)\Big],
\eal
so that 
\be
[\d_\ve,\d_{\o}]\mathscr{W}=0.
\ee

\begin{flushleft}
\begin{tabular}{l}
\hline\hline
\raisebox{0.1cm}{\mbox{$[\d_{\ve},\d_{\ve'}]\mathscr{W}=(\d_\th+\d_\o)\mathscr{W}$ with $\th=-\frac12(\lbar\ve'\g^\m\ve) A_\m$, $\o^I=-\frac12(\lbar\ve'\g^\m\ve) a^I_\m$:\rule{0cm}{.5cm}\hspace{4.8cm}}}\\
\hline\hline
\end{tabular}
\end{flushleft}

Acting with two successive supersymmetry transformations on the generating functional gives 
\bal
\d_{\ve'} \d_{\ve}\mathscr{W}=&\;\k^{(7)}_I\int d^4x\;e\;\Big[A_\m \lbar\ve\big(\g^5\g^\m\d_{\ve'}\l^{I}+i\e^{\m\n\r\s}a_\n^{I}\g_\r\d_{\ve'}\j_\s\big)\NO\\
&\;\hspace{3.0cm}+\frac{i}{2} a_\m^I\lbar\ve\g^\m\g^{\r\s}\Big(\cd_\r\d_{\ve'}\j_\s+\frac{i}{2}V^\t\g_\r\g_\t\g^5\d_{\ve'}\j_\s\Big)\Big]\NO\\
=&\;\k^{(7)}_I\int d^4x\;e\;A_\m \lbar\ve\Big[-\frac14\g^5\g^\m\big(\g^{\r\s} f^I_{\r\s}+\g^5D^I\big)\ve'+i\e^{\m\n\r\s}a_\n^{I}\g_\r\Big(\cd_\s\ve'+\frac{i}{2}V^\k\g_\s\g_\k\g^5\ve'\Big)\Big]\NO\\
&\;+\frac{i}{2}\k^{(7)}_I\int d^4x\;e\; a_\m^I\lbar\ve\g^\m\g^{\r\s}\Big(\cd_\r+\frac{i}{2}V^\t\g_\r\g_\t\g^5\Big)\Big(\cd_\s+\frac{i}{2}V^\k\g_\s\g_\k\g^5\Big)\ve'.
\eal
Using the identities 
\bal
&\Big[\Big(\cd_\r+\frac{i}{2}V^\k\g_\r\g_\k\g^5\Big),\Big(\cd_\s+\frac{i}{2}V^\l\g_\s\g_\l\g^5\Big)\Big]\ve=\\
&\Big(\frac14 R_{\r\s\k\l}\g^{\k\l}+i\g^5G_{\r\s}\Big)\ve+\frac{i}{2}\big(\de_\r V_\k\g_\s-\de_\s V_\k\g_\r\big)\g^\k\g^5\ve-\frac14V^\k V^\l(\g_\r\g_\k\g_\s\g_\l-\g_\s\g_\k\g_\r\g_\l)\ve,\NO
\eal
and 
\bal\label{commutator-identities}
&\lbar\ve'\g^\m\g^{\r\s}\g^5\ve-\lbar\ve\g^\m\g^{\r\s}\g^5\ve'=\lbar\ve'\{\g^\m,\g^{\r\s}\}\g^5\ve=2i\e^{\m\n\r\s}\lbar\ve'\g_\n\ve,\NO\\
&\lbar\ve'\g_\r\cd_\s\ve-\lbar\ve\g_\r\cd_\s\ve'=\de_\s(\lbar\ve'\g_\r\ve),\NO\\
&\lbar\ve'\g^\m\g^\r\g^\s\g^5\ve-\lbar\ve\g^\m\g^\r\g^\s\g^5\ve'=2i\e^{\m\n\r\s}\lbar\ve'\g_\n\ve,
\eal
we therefore obtain 
\bal
[\d_{\ve},\d_{\ve'}]\mathscr{W}=&\;\k^{(7)}_I\int d^4x\;e\;A_\m \big(i\e^{\m\n\r\s}f^I_{\r\s}\x_\n+D^I\x^\m+2i\e^{\m\n\r\s}a_\n^{I}\de_\s\x_\r+8iV^\k \d^{[\m}_\k\d^{\n]}_\l a_\n^{I}\x^\l\big)\NO\\
&\;+\frac{i}{2}\k^{(7)}_I\int d^4x\;e\;a_\m^I\Big(-R\x^\m-6V^2\x^\m-2\e^{\m\n\r\s}\x_\n G_{\r\s}-6\e^{\m\n\r\s}\x_\n\pa_\r V_\s\Big)\NO\\
=&\;-\k^{(7)}_I\int d^4x\;e\; \th\big(D^I-i\e^{\m\n\r\s}\pa_\n B_{\r\s} a_\m^I+\co(\l\j)\big)\NO\\
&\;-\k^{(7)}_I\int d^4x\;e\;\o^I\Big(-\frac{i}{2}R-3iV_\m V^\m+i\e^{\m\n\r\s}\pa_\n B_{\r\s} A_\m+\co(\j^2)\Big),
\eal
where again
\be
\x^\m=\frac12(\lbar\ve'\g^\m\ve),\qquad \th=-\frac12(\lbar\ve'\g^\m\ve) A_\m,\qquad \o^I=-\frac12(\lbar\ve'\g^\m\ve) a^I_\m.
\ee

\subsection{$\k^{(8)}_{[IJ]}$ cocycle}

\begin{flushleft}
\begin{tabular}{l}
\hline\hline
\raisebox{0.1cm}{\mbox{$[\d_\ve,\d_\th]\mathscr{W}=0$:\rule{0cm}{.5cm}\hspace{13.8cm}}}\\
\hline\hline
\end{tabular}
\end{flushleft}

This commutator is trivially satisfied for the cocycle $\k^{(8)}_{[IJ]}$:
\bal
\d_{\ve}\d_{\th} \mathscr{W}=&\;-\d_{\ve}\int d^4x\;e\left.\th\ca_{R}\right|_{\k^{(8)}_{[IJ]}}=0,\\\NO\\
\d_{\th} \d_{\ve}\mathscr{W}=&\;-\k^{(8)}_{[IJ]}\d_{\th}\int d^4x\;e\;\lbar\ve\ca_Q^{(8)[IJ]}=0.
\eal
Hence,
\be
[\d_\ve,\d_\th]\mathscr{W}=0.
\ee

\begin{flushleft}
\begin{tabular}{l}
\hline\hline
\raisebox{0.1cm}{\mbox{$[\d_\ve,\d_{\o}]\mathscr{W}=0$:\rule{0cm}{.5cm}\hspace{13.7cm}}}\\
\hline\hline
\end{tabular}
\end{flushleft}

The calculation in this case is identical to that for the commutator $[\d_\ve,\d_\th]\mathscr{W}=0$ of the $\k^{(7)}_I$ cocycle above. Following the same steps we have 
\bal
\d_{\ve}\d_{\o} \mathscr{W}=&\;-\k^{(8)}_{[IJ]}\d_\ve\int d^4x\;e\;\o^I\big(D^J-i\e^{\m\n\r\s}a^J_\m \pa_\n B_{\r\s}+\lbar\l^J\g^5\g^\m\j_\m\big)\NO\\
=&\;-\k^{(8)}_{[IJ]}\int d^4x\;e\;\o^I\Big[\lbar\ve\g^5\g^\m\Big(\cd_\m\l^J+\frac14(\g^{\r\s} f^J_{\r\s}+\g^5D^J)\j_\m\Big)\NO\\
&\;-\frac{i}{2}\e^{\m\n\r\s}\pa_\n B_{\r\s}\lbar\ve\g_\m\l^J-i\e^{\m\n\r\s}\pa_\n(\lbar\ve\g_\r\j_\s)a^J_\m+\frac12\lbar\ve\g^\m\j_\m D^J\NO\\
&\;+\lbar\l^J\g^5\g^\m\Big(\cd_\m\ve+\frac{i}{2}V^\k\g_\m\g_\k\g^5\ve\Big)-\frac{1}{4}\lbar\j_\m\g^5\g^\m\big(\g^{\r\s}f_{\r\s}+\g^5 D^J\big)\ve\Big]\NO\\
=&\;\k^{(8)}_{[IJ]}\int d^4x\;e\;\pa_\m\o^I \lbar\ve\big(\g^5\g^\m\l^{J}+i\e^{\m\n\r\s}a_\n^{J}\g_\r\j_\s\big),\\
\d_{\o} \d_{\ve}\mathscr{W}=&\;-\k^{(8)}_{[IJ]}\d_{\o}\int d^4x\;e\;\lbar\ve\ca_Q^{(8)[IJ]}\NO\\
=&\;\k^{(8)}_{[IJ]}\int d^4x\;e\;\pa_\m\o^I \lbar\ve\big(\g^5\g^\m\l^{J}+i\e^{\m\n\r\s}a_\n^{J}\g_\r\j_\s\big),
\eal
and so 
\be
[\d_\ve,\d_{\o}]\mathscr{W}=0.
\ee

\begin{flushleft}
\begin{tabular}{l}
\hline\hline
\raisebox{0.1cm}{\mbox{$[\d_{\ve},\d_{\ve'}]\mathscr{W}=(\d_\th+\d_\o)\mathscr{W}$ with $\th=-\frac12(\lbar\ve'\g^\m\ve) A_\m$, $\o^I=-\frac12(\lbar\ve'\g^\m\ve) a^I_\m$:\rule{0cm}{.5cm}\hspace{4.8cm}}}\\
\hline\hline
\end{tabular}
\end{flushleft}

This commutator is also a special case of the corresponding one for the $\k^{(7)}_I$ cocycle: 
\bal
\d_{\ve'} \d_{\ve}\mathscr{W}=&\;\k^{(8)}_{[IJ]}\int d^4x\;e\;a_\m^I \lbar\ve\Big(\g^5\g^\m\d_{\ve'}\l^{J}+\frac{i}{2}\e^{\m\n\r\s}a_\n^{J}\g_\r\d_{\ve'}\j_\s\Big)\\
=&\;\k^{(8)}_{[IJ]}\int d^4x\;e\;a_\m^I \lbar\ve\Big[-\frac14\g^5\g^\m\big(\g^{\r\s} f^J_{\r\s}+\g^5D^J\big)\ve'+\frac{i}{2}\e^{\m\n\r\s}a_\n^{J}\g_\r\Big(\cd_\s\ve'+\frac{i}{2}V^\k\g_\s\g_\k\g^5\ve'\Big)\Big].\NO
\eal
Using the identities \eqref{commutator-identities} we obtain
\bal
[\d_{\ve},\d_{\ve'}]\mathscr{W}=&\;\k^{(8)}_{[IJ]}\int d^4x\;e\;a_\m^I \big(i\e^{\m\n\r\s}f^J_{\r\s}\x_\n+D^J\x^\m+i\e^{\m\n\r\s}a_\n^{J}\de_\s\x_\r+4iV^\k \d^{[\m}_\k\d^{\n]}_\l a_\n^{J}\x^\l\big)\NO\\
=&\;-\k^{(8)}_{[IJ]}\int d^4x\;e\; \o^{I}\big(D^J-i\e^{\m\n\r\s}\pa_\n B_{\r\s} a_\m^J+\co(\l\j)\big),
\eal
as required by the Wess-Zumino consistency conditions.


\bibliographystyle{jhepcap}
\bibliography{NManomalies,CSanomalies}

\end{document}